\documentclass[pre,eps,aps,reprint,twocolumn,superscriptaddress]{revtex4-2}
\usepackage{graphicx}
\usepackage{amsmath}
\usepackage{amssymb}
\usepackage{color}
\usepackage[latin2]{inputenc}
\def\slaninafigdir{.}
\begin{document}
\title{%
Random cubic graph embedded in a hypercube: Entanglement spectrum and many-body localization
}%%
\author{%
Franti\v{s}ek Slanina%
}%
\affiliation{%
Institute of Physics,
 Czech Academy of Sciences,
 Na~Slovance~2, CZ-18200~Praha,
Czech Republic%
}%
\email{
slanina@fzu.cz
}%
\begin{abstract}
The schematic model of interacting spins is introduced, which combines
the symmetry of  hypercube with the simplicity of random regular graph
with degree three, i.e. the random cubic graph. We study the localization
transition in this model, which shares essential characteristics with
the systems exhibiting many-body localization. Namely, we investigate the
transition in terms of the entanglement entropy and entanglement
spectrum. We also show that the most significant indicator of the
localization transition is the failure of eigenstate thermalization
hypothesis, when the distribution of matrix elements of local
operators changes from Gasussian to bimodal. It also provides good
estimate for the critical 
disorder strength.
\end{abstract}
%
%\date{\today}%
%
\maketitle%
\section{Introduction}

The hypothesis that a physical system left in isolation eventually
thermalizes to uniquely determined equilibrium state is the basic principle of all
thermodynamics \cite{lan_lif_80}. However, experimental techniques are
currently available to
prepare systems which violate this principle on large enough time
scales to  make the breakdown of thermalization physically
relevant. It was observed  notably in strongly interacting quantum
systems well isolated from their
environment. The phenomenon acquired the name many-body localization
and it was reported in many experiments
\cite{kon_mcg_xu_dem_15,schreiber_etal_15,alv_sut_kai_15,kau_tai_luk_ris_sch_pre_gre_16,choi_etal_16,bor_lus_hod_sch_blo_sch_16,smi_lee_ric_ney_hes_hau_hey_hus_mon_16,lus_bor_sch_ale_alt_sch_blo_17,bor_lus_sch_gop_kna_sch_blo_17,bor_lus_sch_kna_blo_17,wei_ram_cap_18,xu_che_etal_18,rub_cho_zei_hol_rui_blo_gro_19,koh_sch_li_lus_das_blo_aid_19,ris_luk_sch_kim_tai_leo_gre_19,luk_ris_sch_tai_kau_cho_khe_leo_gre_19,brydges_etal_20,guo_etal_21,gon_mor_zha_etal_21,fil_gon_sag_etal_22}.
The technique used most often involves atoms trapped in an optical
speckle, which forms a quasiperiodic potential, like in the
Aubry-Andr\'e model \cite{aub_and_80}, where single-particle
localization was proved. The trapped atoms can be both fermions
\cite{kon_mcg_xu_dem_15,schreiber_etal_15,lus_bor_sch_ale_alt_sch_blo_17,bor_lus_hod_sch_blo_sch_16,rub_cho_zei_hol_rui_blo_gro_19}
or bosons
\cite{kau_tai_luk_ris_sch_pre_gre_16,choi_etal_16,rub_cho_zei_hol_rui_blo_gro_19,ris_luk_sch_kim_tai_leo_gre_19,luk_ris_sch_tai_kau_cho_khe_leo_gre_19},
the speckle can be one- as well as two-dimensional
\cite{kau_tai_luk_ris_sch_pre_gre_16,bor_lus_sch_gop_kna_sch_blo_17,koh_sch_li_lus_das_blo_aid_19}
and can be also time-periodic, thus forming a Floquet system  \cite{bor_lus_sch_kna_blo_17}.
Other experimental setups are based on
nuclear spins \cite{alv_sut_kai_15,wei_ram_cap_18}
ions \cite{brydges_etal_20} and
quantum simulators and superconducting qubits
\cite{smi_lee_ric_ney_hes_hau_hey_hus_mon_16,xu_che_etal_18,guo_etal_21,gon_mor_zha_etal_21,fil_gon_sag_etal_22}.
Most of the experiments rely on the demonstration of persistence of
initial conditions, thus manifesting non-ergodicity. From the point of view of our work the most
stimulating experiments aim at measuring entanglement entropy
\cite{luk_ris_sch_tai_kau_cho_khe_leo_gre_19,brydges_etal_20}
which is notoriously very difficult
\cite{isl_ma_pre_tai_luk_ris_gre_15}. %(this work is not on MBL)

{
The study of the localization phenomenon, pioneered by Anderson
\cite{anderson_58}, concentrated for a long time on non-interacting
systems, where the existence of localized phase was rigorously proved
in various
geometries, starting from the Bethe lattice \cite{ku_sou_83}.
However, it might be interesting to recall that already in the
original Anderson'a paper  \cite{anderson_58} there is a hint that
localization might put in question the assumed approach to thermal
equilibrium in interacting spins systems perfectly  isolated from the
external heat bath, which is analogous to the situation observed in recent
experiments mentioned above.
}

{
Historically, the debate about the localization in many-particle
interacting systems was long
open,} with general belief tending to the opinion that however weak
interaction always destroys localization and the many-body systems can
at most bear some traces of localized character that would
hypothetically exist if the interaction was turned off.

Such mood changed after a series of papers
\cite{gor_mir_pol_05,bas_ale_alt_06} which presented convincing
arguments based on diagrammatic approaches, that weak enough
interaction does not disrupt the localized character of many-body
eigenstates. One of the essential ingredients was the similarity of
the topology of the Fock space with a Cayley tree, at least on local
level. The argument is that the localization in the many-particle Fock space
then follows from the well-known localization on a Bethe lattice
\cite{ku_sou_83,ab_an_tho_73}.

These arguments found immediately strong support from exact
diagonalization, time-dependent DMRG and numerical renormalization-group
studies of specific one-dimensional Hamiltonians with on-site disorder
\cite{oga_hus_07,zni_pro_pre_08,pal_hus_10,mon_gar_10a,ber_rei_10}
which was then confirmed and refined in subsequent studies
\cite{bar_pol_moo_12,ser_pap_aba_13,del_sca_13,iye_oga_ref_hus_13,vos_alt_13,kja_bar_pol_14,ser_pap_aba_14,nan_kim_hus_14,and_ens_sir_14,pon_pap_huv_aba_15,mod_muk_15,bar_coh_rei_15,lui_laf_ale_15,bay_lim_she_15,mon_pal_hug_lau_15,li_gan_pix_das_15,ser_pap_aba_15,aga_gop_kna_mul_dem_15,her_san_15,fri_wer_bro_sch_eis_15,ore_mic_ser_sil_19,wei_che_xia_mon_19,mac_ale_laf_19,hop_ors_hei_21}. MBL
phase was then observed also in systems with random interaction,
rather than random potential \cite{bar_rei_sag_16,sie_del_zak_17}, in
deterministic Fibonacci chains \cite{mac_laf_ale_19}, in Floquet
systems \cite{pon_pap_huv_aba_15}, or in absence of
disorder in gauge-invariant systems \cite{bre_dal_hey_sca_18}.

The rigorous proof of the existence of MBL state is
still lacking, but current mathematical results come quite close to
it. Indeed, MBL was proved in spin chains based on a very plausible
(yet unproven) assumption on the eigenenergy spectrum
\cite{imbrie_16,ros_mul_sca_15}. Further rigorous results concern spin systems
which can be mapped on free fermions via Jordan-Wigner (JW) transform
\cite{sim_sto_15}. This result is sometimes erroneously considered
obvious, but it is far from that due to non-locality of the JW
transform. There are also rigorous results on localization in spin
chains infinitesimally close to ground state \cite{stolz_20} and on
interacting systems with arbitrary (but finite in thermodynamic limit)
number of particles \cite{chu_suh_09,ekanga_21}.

The phenomenology of MBL state is usually related to the emergence of
the full set of local integrals of motion (LIOM). It is supposed that
the Hamiltonian of the system can be written as a sum of products of
these LIOMs, with coefficients which decay exponentially with
distance \cite{hus_nan_oga_14,ser_pap_aba_13a,cha_kim_vid_aba_15,rad_ort_16,imbrie_16,ros_mul_sca_15,bri_aba_vid_pap_16,imb_ros_sca_17,rad_ort_som_17}. On one side, this approach is a basis of
proofs of the existence of MBL state \cite{imbrie_16,ros_mul_sca_15},
on the other it
serves as a starting point for investigation of the stability of MBL
state with respect to small heat baths \cite{gop_nan_14,nan_gop_hus_14,lui_huv_der_17}, Griffiths effects
\cite{aga_gop_kna_mul_dem_15,pot_vas_par_15,vos_hus_alt_15,gop_mul_khe_kna_dem_hus_15,gop_aga_dem_hus_kna_16} and avalanches \cite{der_huv_17,thi_huv_mul_der_18,mor_col_khe_lui_hus_22}. However, it is not yet
clear how the scheme of LIOMs works in systems with many-particle
mobility edge. In fact, the very existence of MBL state with a
mobility edge was questioned \cite{der_huv_mul_sch_16}, contrary to
what the exact
diagonalization studies suggest
\cite{lui_laf_ale_15,bay_lim_she_15,mon_pal_hug_lau_15,ore_mic_ser_sil_19,wei_che_xia_mon_19}.

The essence of the transition to MBL state is abrupt change in the
character of the eigenvectors of the Hamiltonian. This manifests
itself in broken ergodicity which can be measured by several witness
indicators. The first witness is the level statistics, which
changes from Wigner-Dyson to Poisson \cite{oga_hus_07} and it was used
in most of the exact diagonalization studies of MBL. The second is the
entanglement entropy, where volume law is replaced by area law
\cite{bau_nay_13}. In ergodic phase, entanglement entropy grows
linearly in the asymptotics of large sizes. For finite sizes, Page
formula \cite{page_93} is conjectured to hold. On top of it,
exponential corrections follow from the random-matrix theory, as used
in \cite{pie_par_mar_pas_sca_17}. More detailed information is
contained in entanglement spectrum, which has
Mar\v{c}enko-Pastur \cite{mar_pas_67} form in the ergodic phase and qualitatively
different form in MBL phase \cite{yan_cha_ham_muc_15,ser_mic_aba_pap_16,ger_nan_reg_16,ger_reg_nan_17,bui_gri_che_19,pie_par_mar_pas_sca_17}. Moreover,
on the dynamic side, MBL phase is distinguished from both ergodic and
single-particle Anderson localized systems by logarithmic growth of
entanglement entropy (until saturation due to finite size),
when started from a product state
\cite{dec_mon_cal_faz_06,zni_pro_pre_08,nan_kim_hus_14,and_ens_sir_14}.

The third witness is the breakdown of
eigenvector thermalization hypothesis (ETH)
\cite{deutsch_91,srednicki_94,rig_dun_ols_08,reimann_15}.
This can be measured e. g. using the  probability
distribution of the matrix elements of
local operators, which is expected to be Gaussian in ergodic state,
but changes to bimodal form in MBL state
\cite{pal_hus_10,luitz_16,pan_sca_sch_tay_zni_19,goi_eis_kru_19}.

Together, the accumulated evidence leads to widely accepted view
that MBL is a well established phase of matter. On the
other hand, serious doubts do exist about observability of true MBL behavior
with current approaches and available sample sizes
\cite{pan_sca_sch_tay_zni_19,sun_bon_pro_vid_20,sie_del_zak_20,sie_zak_22},
despite efforts in developing new algorithms
\cite{sie_lew_zak_20}.  Further details on various aspects of MBL can
be found in several recent reviews, e. g.
\cite{nan_hus_15,alt_vos_15,aba_pap_17,lui_bar_17,aga_alt_dem_gop_hus_kna_17,altman_18,ale_laf_18,aba_alt_blo_ser_19,gop_par_20}.

In this work we come back to the early stage of the research in MBL
and try to revisit a not fully investigated path. As we already
mentioned, the founding arguments of the basic works
\cite{gor_mir_pol_05,bas_ale_alt_06} relied on the similarity of the
Fock space topology with that of a Bethe lattice, at least on a local
level. In fact, local isomorphism of a graph to a tree is enough to
establish exactly the transition point \cite{ab_an_tho_73} but not
enough to fully establish the nature of eigenstates. The point is that
a finite Bethe lattice, or more properly a Cayley
tree, is fundamentally unphysical due to the fact that the surface sites
comprise finite fraction of the total volume. This has profound
implications on the character of eigenvectors
\cite{bir_sem_tar_10,del_alt_kra_sca_14,son_tik_mir_17}.  There
is a simple way out, though, namely working with (finite) random
regular graphs (RRG) which are locally isomorphic to Cayley trees, but
do not have any boundary by definition.  The localization on RRGs was
studied numerically
\cite{slanina_12b,alt_cue_iof_kra_16,tik_mir_skv_16,det_ber_sca_kha_20}
as well as
analytically \cite{tik_mir_19,tik_mir_19a,tik_mir_21,tik_mir_21a}.
There is a close
relation to Rosenzweig-Porter  random-matrix ensembles, investigated
e. g. on
\cite{ros_por_60,kra_kha_cue_ami_15,kha_kra_alt_iof_20,bir_tar_21,ven_cug_sch_tar_23}.
Note  also the connection
to the model of randomly interacting Majorana fermions
\cite{mon_mic_tez_alt_21}.

Despite basic similarity, there are features of 1D MBL systems which
do not have counterparts in RRG. For example, there are no global
conserved quantities like total magnetization. Therefore,  there is no
obvious way how to define spin transport (connected to  conserved
magnetization) or mass
transport (connected to conserved number of particles).
{That said, we should at the same time stress that the
presence of a conserved quantity, which is needed for sensible
definition of transport coefficients, is not indispensable for MBL
effect itself. As an example we can mention e.g. Floquet systems
\cite{pon_pap_huv_aba_15}.}
The essential
{feature which makes RRG different}
is missing connection to Euclidean space in which spins or
particles would reside. The same holds for spreading of entanglement,
as there is no natural way how to express the Hilbert space of the
system, which has the form of RRG, as a tensor product of two subsystems'
Hilbert spaces.

Therefore, to make connection between features special to MBL and
those of RRG, an additional feature or invention is necessary. A
successful example of this type is the Kosterlitz-Thouless (KT) scaling
documented recently in certain random graphs
\cite{gar_mar_gir_geo_dub_lem_22}. Indeed, two-parametric KT flow was
observed in phenomenological renormalization-group approaches to MBL
\cite{mor_hus_19,dum_gor_par_ser_vas_19,gor_vas_ser_19}. These
calculations rely fundamentally on 1D geometry of the
system and embody the Griffiths phenomena which are believed to drive
the MBL-to-ergodic transition. Neither 1D geometry nor Griffiths
physics is present in random graphs studied in
\cite{gar_mar_gir_geo_dub_lem_22} but still KT flow is observed, which
indicates that perhaps more features considered specific to MBL may be
found in random graphs like RRG, if only correct path is discovered.

{The main purpose of our present work is to amend the
``amorphous'' topology of random regular graphs so that it might
provide insight to further features specific to MBL. To this end,}
we introduce here a class of
random graphs which enable studying entanglement properties and ETH
breaking.  {Thus, it is indispensable that the
geometry of RRG contains additional structure which is isomorphic to
a tensor product. To achieve that,} we introduce random cubic graphs
embedded into a
hypercube of specified dimension. Random cubic graph is  the simplest
type of RRG, fixing the number of neighbors of each vertex to 3. In
our case, the set of vertices coincides with the set of vertices of
the hypercube and the set of edges of the cubic graph is the subset of
edges of the hypercube. In order to construct such a graph, we
introduce a stochastic graph process. Although the algorithm does not
sample the huge set of all allowed graphs uniformly, we believe that
it produces instances which are randomized enough to provide
representative information. In such a model, it is straightforward to
study entanglement entropy and entanglement spectrum. We believe this
model captures what is essential for the MBL transition and contains
just little additional model-specific peculiarities.

\section{Construction of random Hamiltonian}

Consider a system of $L$ $1/2$-spins with yet unspecified
interaction. The Hilbert space of such system is the set of vertices
of $L$-dimensional hypercube
\begin{equation}
\mathcal{H}=\mathcal{V}_\mathrm{hyp}\equiv\{|+1\rangle,|-1\rangle\}^L\;.
\end{equation}
The set of edges of the hypercube consists of pairs of vertices with
Hamming distance $d_H$ exactly one
\begin{equation}
\mathcal{E}_\mathrm{hyp}=\{\{v,v'\}:v,v'\in\mathcal{V}_\mathrm{hyp}\land d_H(v,v')=1\}\;.
\end{equation}
We can define local spin operators acting in this space in an obvious way
\begin{equation}
\sigma_{i\alpha}={\bf 1}_1\otimes\ldots\otimes{\bf 1}_{i-1}
\otimes\sigma_\alpha\otimes{\bf 1}_{i+1}\otimes\ldots\otimes{\bf
1}_L
\label{eq:localspins}
\end{equation}
where $\sigma_\alpha$, $\alpha\in\{x,y,z\}$ are the usual $2\times 2$
Pauli spin matrices.

We want to construct a random Hamiltonian of the system in such a way
that its off-diagonal elements represent an adjacency matrix of a
random graph. We want that this graph is locally similar to a random
regular graph but globally
reflects the hypercube geometry of the Hilbert space. Therefore, we
construct a graph $\mathcal{G}=[\mathcal{V},\mathcal{E}]$ whose
vertices coincide with the vertices of the hypercube
$\mathcal{V}=\mathcal{V}_\mathrm{hyp}$ but the set of edges in only a
subset of the edges of the hypercube, $\mathcal{E} \subset
\mathcal{E}_\mathrm{hyp}$. This subset is chosen randomly with the
only constraint that the degree of all vertices in graph $\mathcal{G}$
is the same, i.e. it is a special instance of a random regular
graph. In this work the fixed degree will be $3$, hence it is a random
cubic graph. Besides the randomness in the structure of the graph, the
Hamiltonian will contain also random diagonal elements. Therefore, we
write the Hamiltonian as a matrix
\begin{equation}
H_{ij}= A_{ij}[\mathcal{G}] +\eta\xi_i\delta_{ij}
\label{eq:hamiltonian}
\end{equation}
where $A_{ij}[\mathcal{G}]$ is the adjacency matrix of the graph
$\mathcal{G}$ described above, $\xi_i$ are i.i.d. random normally
distributed numbers and the positive parameter $\eta$ measures the
strength of diagonal disorder.

\begin{figure}[t]
\includegraphics[scale=0.3]{%
\slaninafigdir/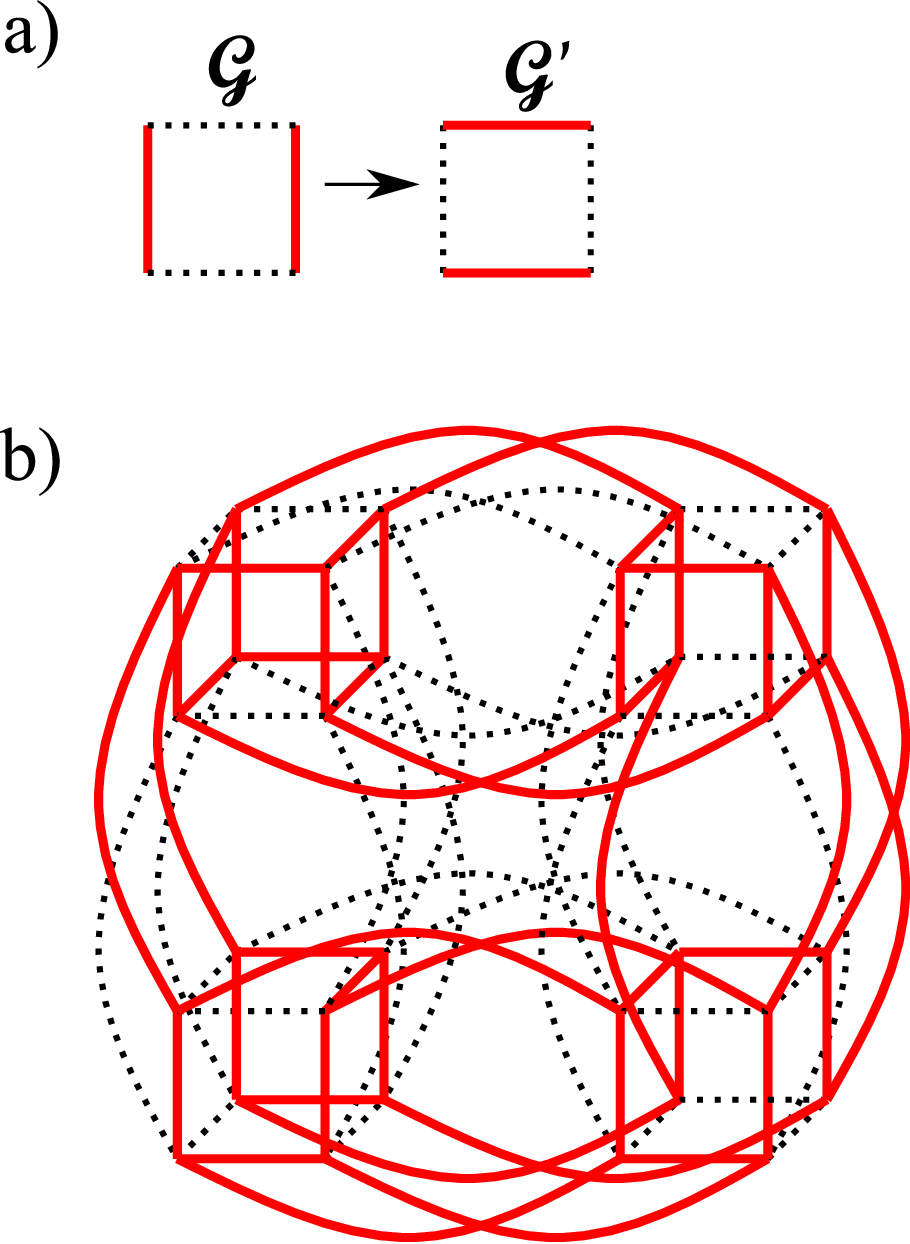}
\caption{In a), scheme of a single rewiring move. Full line represents
an edge contained in the graph $\mathcal{G}$ (before move) or
$\mathcal{G}'$ (after move), dotted line represents an edge
contained in the hypercube, but absent in the graph. In b), an
example of the cubic graph obtained by a random sequence of such moves. In
this case, it is embedded in the five-dimensional hypercube ($L=5$).
}
\label{scheme-move}
\end{figure}

It would be desirable to sample the set of graphs satisfying the above
conditions with uniform probability. Due to computational difficulty
in guaranteeing the uniformity, we suggest an alternative path, which
is algorithmically very simple and we believe that it samples the whole
set of graphs in sufficiently representative way. The algorithm
proceeds in the following sequence of steps. First, choose a starting
graph which is non-random, but cubic
by definition. In our case, we take the first three dimensions of
$L$-dimensional hypercube and place graph edges at all hypercube edges
which go along these dimensions. This way the starting graph contains
$2^{L-3}$ disconnected cubes and is obviously cubic.
Then, perform random rewiring of the graph so that in
each elementary move the degrees of the touched vertices are
preserved. This way, the graph is random but remains cubic all the
time. Each elementary move consists in finding randomly a plaquette, i.e. a
$4$-cycle, on a
hypercube, so that two opposite sides of the plaquette contain edges
of the graph $\mathcal{G}$ and the other two edges are empty. Then, we
rewire the plaquette so that the two edges move to the empty sides of
the plaquette. This way a new, changed graph $\mathcal{G}'$ is
produced.  Within the plaquette, all four vertices have degree one
before as well as after the move and no other vertices are influenced,
so that the new graph $\mathcal{G}'$ has the same degrees of vertices
as the old graph $\mathcal{G}$. The move is illustrated in Fig
\ref{scheme-move}a. After some time, no such plaquettes are present
and the randomization algorithm stops. We also checked that the graphs
produced this way are connected, i.e. they do not contain disjoint
components. An example of a graph created by this algorithm is shown
in Fig. \ref{scheme-move}b. In this case, it is embedded in the
five-dimensional hypercube.

{In order to test whether or not the rewiring
algorithm described above may
induce any artifacts into the results, we compared the properties of
graphs obtained this way with those sampled uniformly for a small
system of dimension $L=5$. The results are shown in detail in the
Appendix and we can conclude that no artifacts are
observed. Therefore, we shall rely on the rewiring algorithm in all
what follows.}

The combination of random graph and random diagonal elements
constitutes the Hamiltonian (\ref{eq:hamiltonian}) which represents
random interaction of the $L$ spins. Contrary to the various models
based on locally interacting Heisenberg spin chains, here we have  no
a priori conserved quantities (like the total spin), which makes a
technical advantage, because the Hilbert space does not contain
disconnected segments and all vertices are statistically equivalent.

\begin{figure}[t]
\includegraphics[scale=0.45]{%
\slaninafigdir/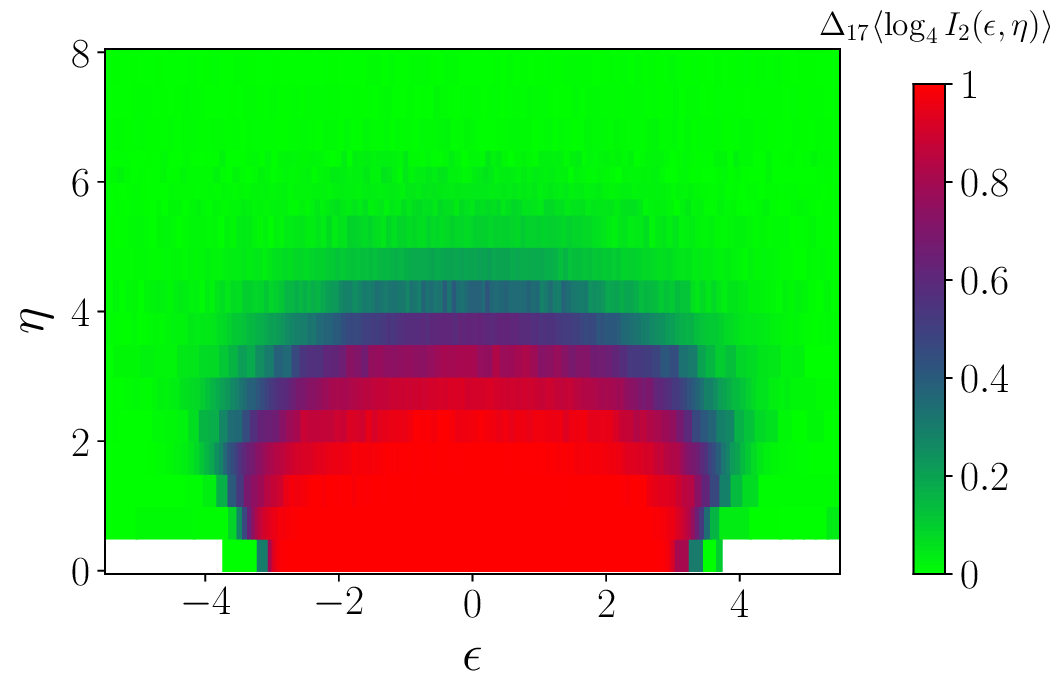}
\caption{Phase diagram of localization in the energy-disorder
plane. Color code shows the value of the difference of log-averaged IPR between
two largest system sizes used. Intermediate color (blue) provides lower
estimate for the position of the mobility edge. White means
absence of data.
}
\label{phasediag-eta-vs-lambda}
\end{figure}

\section{Signatures of localization}

\subsection{Inverse participation ratio}

We shall investigate only static features of localization. From this
point of view, localization can be described as qualitative change in
the dependence of various properties of eigenvectors on system size,
in our case on the dimension $L$ of the hypercube. The most direct of
these properties is the inverse participation ratio (IPR), so we shall investigate
it first. Let us denote
$\psi_n(\epsilon)$ $n$-th element of the eigenvector of the
Hamiltonian (\ref{eq:hamiltonian}) corresponding to eigenvalue
$\epsilon$. We define the moments
\begin{equation}
I_q(\epsilon)=\sum_{n=1}^{2^L}\psi_n^{2q}(\epsilon)\; .
\label{eq:iprandmoments}
\end{equation}
Beyond the obvious normalization $I_1(\epsilon)=1$ the lowest moment is the
IPR, i.e. $I_2(\epsilon)$. For increasing size, $L\to\infty$, it
should scale like
\begin{equation}
\ln I_2(\epsilon)\simeq \kappa_1(\epsilon) L\ln 2
+\kappa_0(\epsilon)\;,
\label{eq:iprscaling}
\end{equation}
where $\kappa_1(\epsilon)=0$ in the localized phase and
$\kappa_1(\epsilon)=1$ in the extended phase.

We investigated the properties of the eigenvectors of
(\ref{eq:hamiltonian}) by full numerical diagonalization of samples
for even $L$ up to size $L=18$. We averaged the logarithm of IPR over
many realizations of the diagonal disorder and random graph and also
over all eigenvectors whose eigenvalues fall
inside a narrow window
$[\epsilon-\Delta\epsilon/2,\epsilon+\Delta\epsilon/2]$. The
width of the window $\Delta\epsilon$ was chosen empirically so that
the effect of stochastic noise was minimized, while the
systematic $\epsilon$-dependence of IPR was well visible. This way we
obtained the numerical average $\langle\ln I_2(\epsilon,\eta)\rangle$ as a
function of energy $\epsilon$, the disorder strength $\eta$ and system
size $L$.

\begin{figure}[t]
\includegraphics[scale=0.45]{%
\slaninafigdir/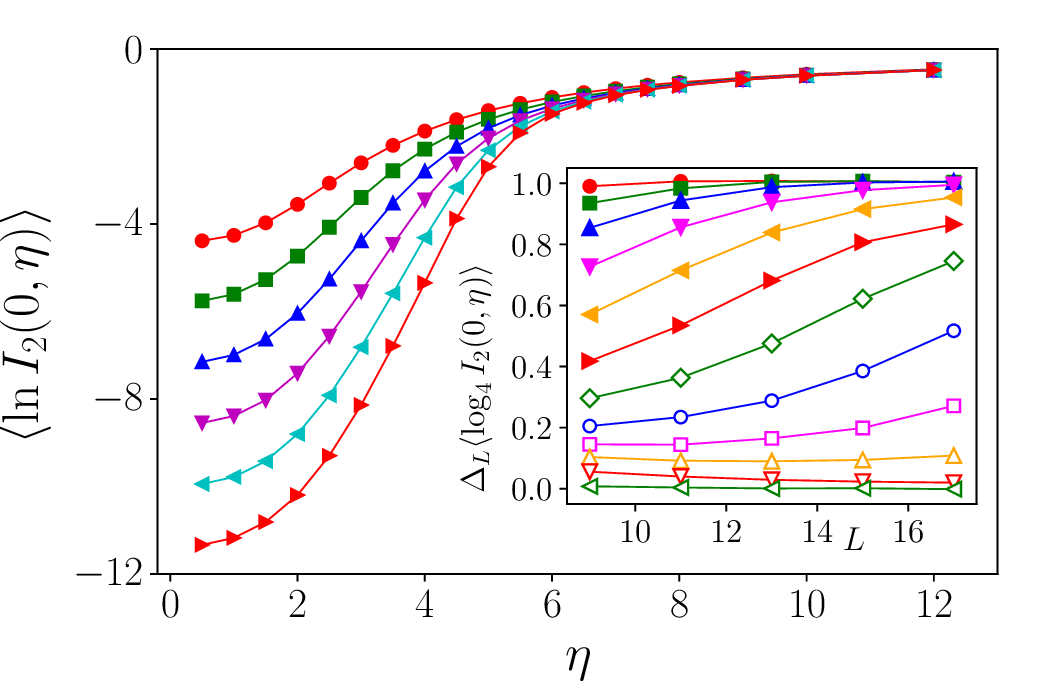}
\caption{Dependence of the inverse participation ratio on the
disorder strength at the center of the
band, $\epsilon=0$. Different symbols indicate size $L=8$ ({\Large $\bullet$}), $10$ ($\blacksquare$), $12$ ($\blacktriangle$),
$14$ ($\blacktriangledown$), $16$ ($\blacktriangleleft$), $18$ ($\blacktriangleright$).
In the inset, the size dependence of the difference in log-averaged
IPR. The values of disorder strength are
$\eta=0.5$ ({\Large $\bullet$}),
$1.5$ ($\blacksquare$),
$2  $ ($\blacktriangle$),
$2.5$ ($\blacktriangledown$),
$3  $ ($\blacktriangleleft$),
$3.5$ ($\blacktriangleright$),
$4  $ ({\Large $\diamond$}),
$4.5$ ({\Large $\circ$}),
$5  $ ($\Box$),
$5.5$ ($\bigtriangleup$),
$6.5$ ($\bigtriangledown$),
$12 $ ({\Large $\triangleleft$}).
}
\label{ipr-stred}
\end{figure}
\begin{figure}[t]
\includegraphics[scale=0.45]{%
\slaninafigdir/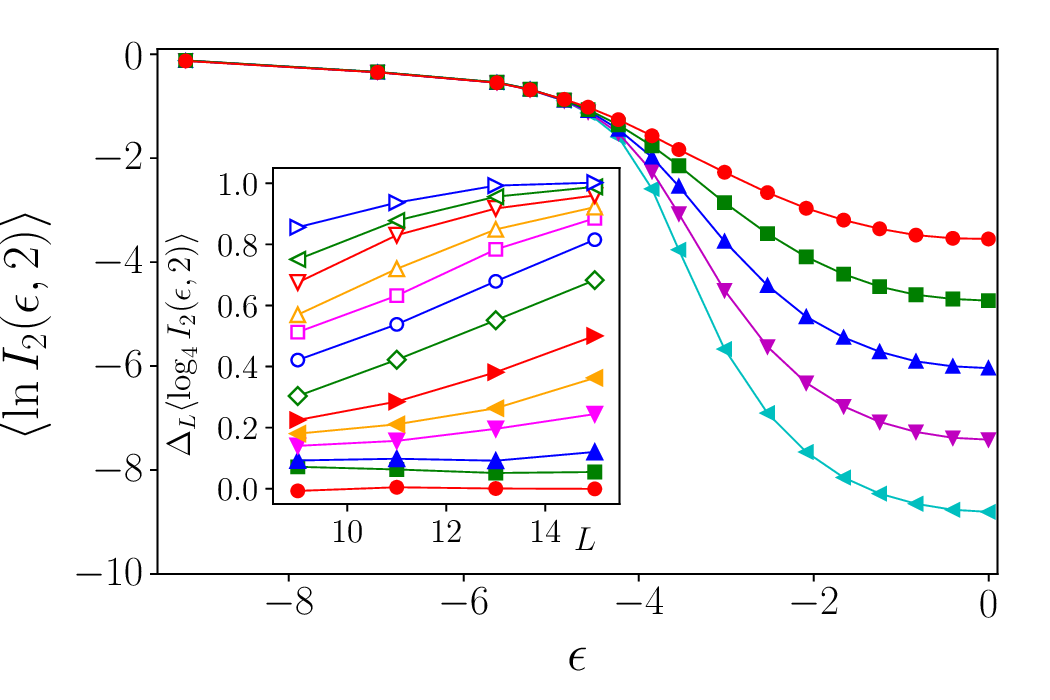}
\caption{Dependence of the inverse participation ratio on the
energy, for disorder strength fixed at $\eta=2$.
Different symbols indicate size $L=8$ ({\Large $\bullet$}), $10$ ($\blacksquare$), $12$ ($\blacktriangle$),
$14$ ($\blacktriangledown$), $16$ ($\blacktriangleleft$).
In the inset, the size dependence of the difference in log-averaged
IPR. The values of the energy are
$\epsilon=-5.620893$ ({\Large $\bullet$}),
$-4.232939$ ($\blacksquare$),
$-4.069776$ ($\blacktriangle$),
$-3.848635$ ($\blacktriangledown$),
$-3.714343$ ($\blacktriangleleft$),
$-3.542069$ ($\blacktriangleright$),
$-3.275039$ ({\Large $\diamond$}),
$-3.019555$ ({\Large $\circ$}),
$-2.764402$ ($\Box$),
$-2.528745$ ($\bigtriangleup$),
$-2.086117$ ($\bigtriangledown$),
$-1.658138$ ({\Large $\triangleleft$}),
$-0.003667$ ({\Large $\triangleright$}).
}
\label{ipr-eta2}
\end{figure}

The scaling (\ref{eq:iprscaling}) holds in the asymptotic regime
$L\to\infty$. For finite sizes we have at our disposal,
we introduce the differences
\begin{equation}
\begin{split}
\Delta_{\overline{L}}\langle & \log_4 I_2(\epsilon,\eta)\rangle=\\
&  \frac{\langle\ln I_2(\epsilon,\eta)\rangle_{L=\overline{L}-1}-
\langle\ln I_2(\epsilon,\eta)\rangle_{L=\overline{L}+1}}{2\ln 2}
\end{split}
\end{equation}
and estimate the factor
$\kappa_1$ by difference between two largest values of $L$ available
i.e. $L=16$ and $L=18$
\begin{equation}
\kappa_1(\epsilon)\simeq\Delta_{17}\langle\log_4 I_2(\epsilon,\eta)\rangle\;.
\end{equation}
In the plane energy-disorder, the quantity
$\Delta_{17}\langle\log_4 I_2(\epsilon,\eta)\rangle$
approximates the phase diagram, separating
the extended and localized regime. We show this approximate phase
diagram in Fig. \ref{phasediag-eta-vs-lambda}. We can see that for
small disorder $\eta$ the central part of the spectrum contains
extended states and the localized states are in the tails. Therefore,
the system is characterized by the presence of the mobility edge.
However for disorder larger than the critical value $\eta_c$ all
eigenvectors
are localized.

For the purpose of establishing the value of the critical disorder the phase
diagram  in Fig. \ref{phasediag-eta-vs-lambda} is not sufficient,
because it tells us little on the dependence on $L$. We
show in Fig. \ref{ipr-stred} the section for the energy fixed at the
center of the band, $\epsilon=0$, and
the size-dependence is explicitly shown. We can see that the
log-averaged IPR decreases with $L$ in the extended regime, while it
remains constant in the localized regime. Even more detailed picture
shows the inset in  Fig. \ref{ipr-stred} where we plot the flow of the
difference $\Delta_{{L}}\langle  \log_4 I_2(\epsilon,\eta)\rangle$
with increasing $L$. The asymptotic values $0$ and $1$ of this
difference correspond to localized and extended phase,
respectively. The flow lines go clearly to $1$ for small enough
$\eta$, but for intermediate $\eta$ they first seem to decrease toward
$0$, but then turn up and tend to $1$. These flow lines and their
corresponding $\eta$ should be also considered as belonging to
extended phase, despite the low value of the difference
$\Delta_{{L}}\langle  \log_4 I_2(\epsilon,\eta)\rangle$.
So, from this flow diagram we
can guess the lower estimate for the critical disorder $6<\eta_c$.

Similarly we can analyze the section of the phase diagram for fixed
disorder strength. We show in Fig. \ref{ipr-eta2} one such section,
the dependence of log-averaged IPR on energy, for $\eta=2$. For
simplicity we show just one half of the dependence, for
$\epsilon<0$. It is sufficient due to the
$\epsilon\to -\epsilon$ symmetry of the averaged IPR. For
$\eta<\eta_c$ the spectrum contains two symmetric mobility edges at
energies $-\epsilon_m$ and $\epsilon_m$; the states
for $|\epsilon|<\epsilon_m$ are extended, while the states in the
tails $|\epsilon|>\epsilon_m$ are localized. Again, we show in the
inset of Fig.  \ref{ipr-eta2} a flow diagram, which shows how the
difference in log-averaged IPR tends to $0$ for localized states and
to $1$ for extended states. Again, for intermediate energies we can
see how the flow line first seem to tend to $0$, but then turns up
toward $1$. Therefore, the situation is similar as with the estimate
of the critical
disorder $\eta_c$. Here also, from the flow diagram we can get a
lower estimate for the mobility edge, in this example for fixed
$\eta=2$ this estimate would be $\epsilon_m>4.0$

\begin{figure}[t]
\includegraphics[scale=0.45]{%
\slaninafigdir/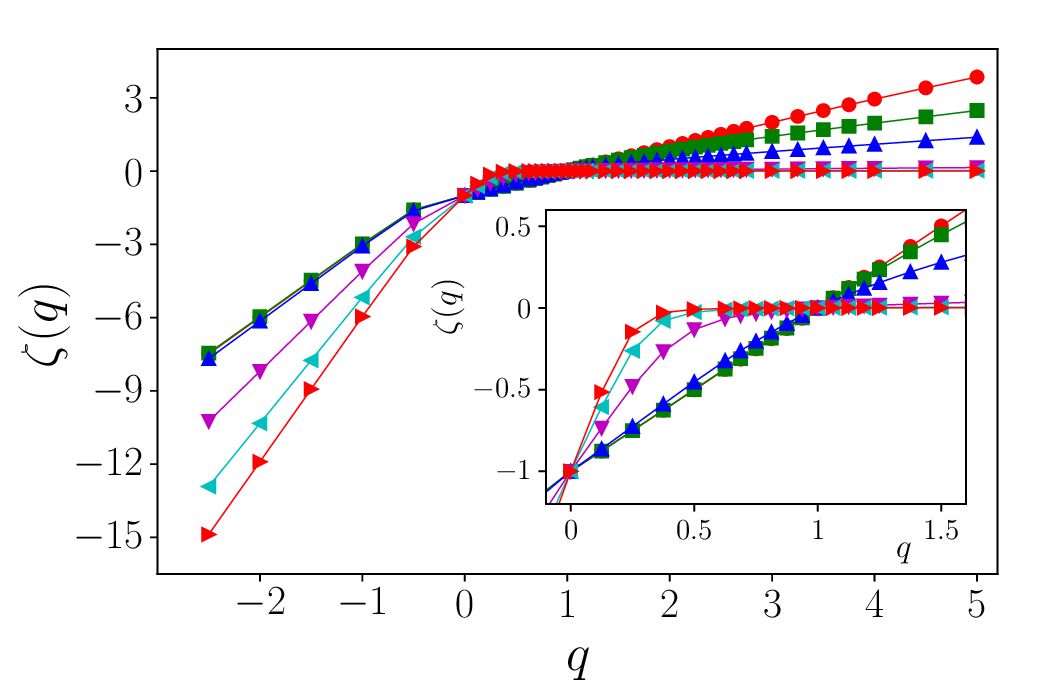}
\caption{Exponents for moments $I_q$, obtained as in
Eq. (\ref{eq:defexponents}) for
$L\le18$.
The eigenvectors are taken at the center of the band,
$\epsilon=0$. The disorder strength is
$\eta=1$ ({\Large $\bullet$}),
$3$ ($\blacksquare$),
$4$ ($\blacktriangle$),
$6$ ($\blacktriangledown$),
$8$ ($\blacktriangleleft$),
$10$ ($\blacktriangleright$). In the inset, the detail of the
dame data is shown.
}
\label{exp-for-mul-sp}
\end{figure}
\begin{figure}[t]
\includegraphics[scale=0.45]{%
\slaninafigdir/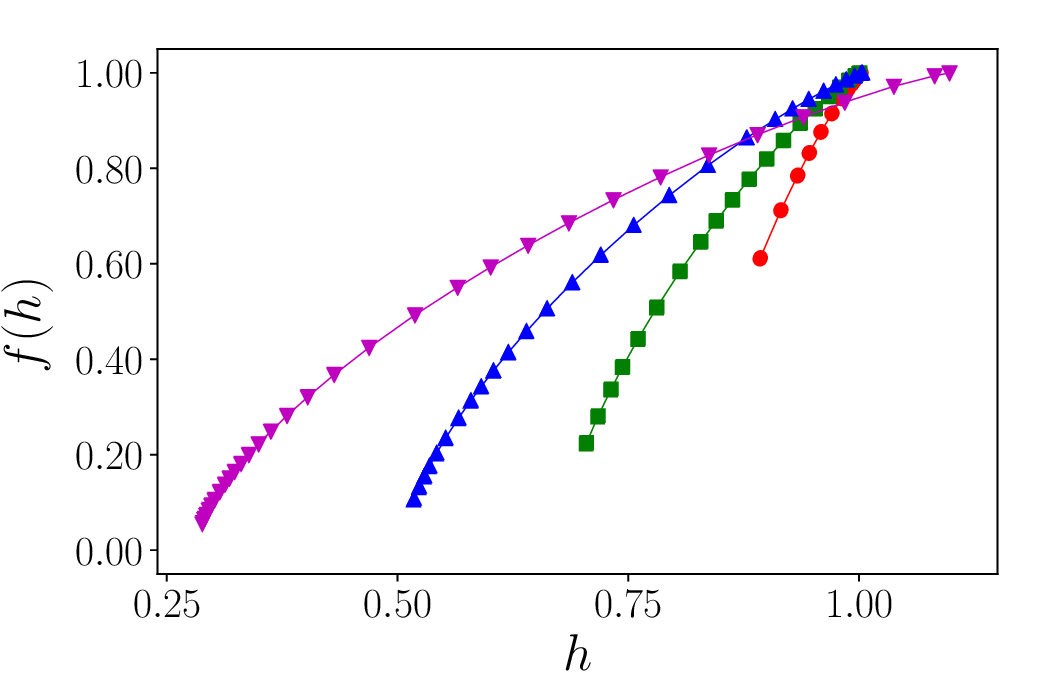}
\caption{Multifractal spectrum obtained by numerical inversion of the
Legendre transform (\ref{eq:legendre}), for the data shown in
Fig. \ref{exp-for-mul-sp}. The disorder strength is
$\eta=1$ ({\Large $\bullet$}),
$2$ ($\blacksquare$),
$3$ ($\blacktriangle$),
$4$ ($\blacktriangledown$).
}
\label{mul-sp-del}
\end{figure}
\begin{figure}[t]
\includegraphics[scale=0.45]{%
\slaninafigdir/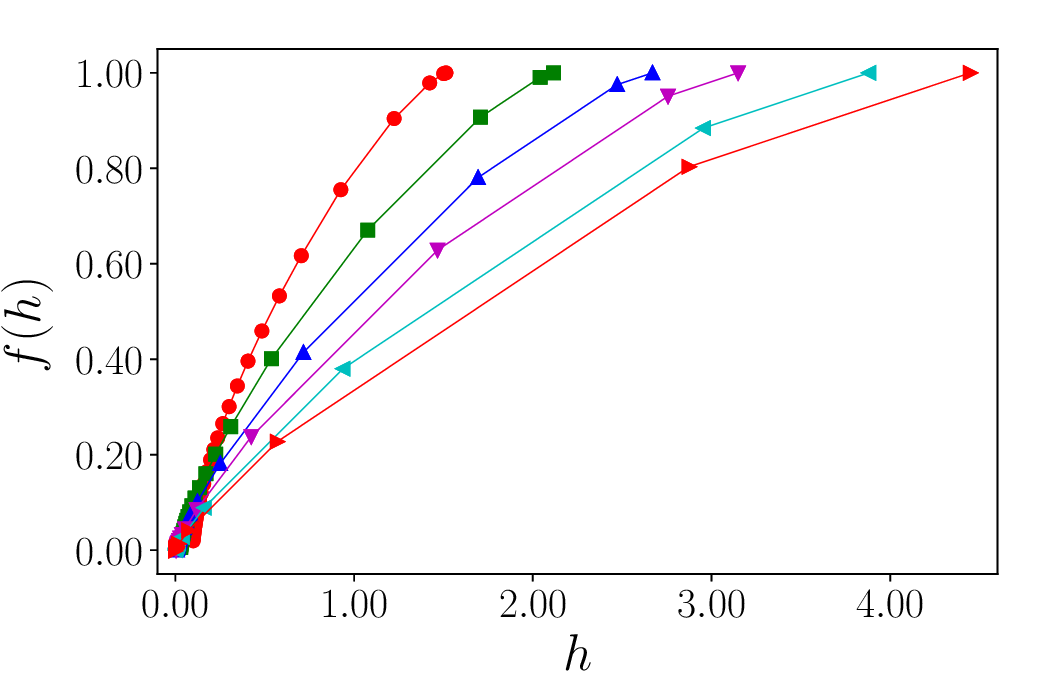}
\caption{Multifractal spectrum obtained by numerical inversion of the
Legendre transform (\ref{eq:legendre}), for the data shown in
Fig. \ref{exp-for-mul-sp}. The disorder strength is
$\eta=5$ ({\Large $\bullet$}),
$6$ ($\blacksquare$),
$7$ ($\blacktriangle$),
$8$ ($\blacktriangledown$),
$10$ ($\blacktriangleleft$),
$12$ ($\blacktriangleright$).
}
\label{mul-sp-loc}
\end{figure}

\subsection{Multifractality}

We extract the multifractal properties of the eigenvectors from the
size dependence of the moments (\ref{eq:iprandmoments}). As the
dimension of the Hilbert space is $N=2^L$, we expect for large $L$ the
asymptotic behavior
\begin{equation}
\langle\ln I_q\rangle \simeq -\zeta(q)L\ln 2\;,\qquad L\to\infty\;.
\label{eq:defexponents}
\end{equation}
Obviously, $\zeta(1)=0$ and $\zeta(0)=-1$. We can express the
exponents $\zeta(q)$ as
Legendre transform of the multifractal spectrum $f(h)$  \cite{slanina_17}
\begin{equation}
\zeta(q)=\min_h(qh-f(h))\;.
\label{eq:legendre}
\end{equation}
In general, the function $f(h)$ is defined on a set of non-negative
numbers $h\in\mathcal{M}$, where $\mathcal{M}$ can contain isolated
points or continuous intervals or both. The minimum in
(\ref{eq:legendre}) is taken over the whole definition domain $\mathcal{M}$.
For example, homogeneous
eigenvector has $\mathcal{M}=\{1\}$ and
$f(1)=1$. Localized states are characterized by the presence of the
point $0\in\mathcal{M}$. For the eigenvectors of a matrix from GOE,
one can find that $\mathcal{M}=\{1,3\}$, with $f(1)=1$ and $f(3)=0$.
Non-trivial multifractal behavior occurs when
the multifractal spectrum
$f(h)$ contains a continuous part \cite{slanina_17}.

We calculated the log-averaged moments for the center of the band,
$\epsilon=0$ and extracted the exponents from the $L$-dependence. The
result is shown in Fig. \ref{exp-for-mul-sp}. From the data we can
make several observations. First, for both $q\to\infty$ and
$q\to -\infty$ the exponent $\zeta(q)$ approaches a linear
function. In fact, if $\zeta(q)$ were a piecewise-linear function,
then each linear piece would correspond to an isolated point in the
set $\mathcal{M}$. The position of the point is given by the slope of
the linear piece. This implies that genuine multifractality occurs
when $\zeta(q)$ contains a piece with continuously changing tangent.
We can see in the detail shown in inset in Fig. \ref{exp-for-mul-sp}
that this is the case of curves corresponding to disorder strengths
$\eta=6$, $8$, $10$. These values are therefore candidates for genuine
multifractal behavior.

To see it better, we inverted numerically the
Legendre transform (\ref{eq:legendre}) in order to find the function
$f(h)$. We show the results in Figs. \ref{mul-sp-del} and
\ref{mul-sp-loc}. Each point in these figures is a pair $(h,f(h))$
which is compatible with the data shown in Fig.
\ref{exp-for-mul-sp}. In Fig.  \ref{mul-sp-del} we show the
multifractal spectrum for weak disorder, below the expected
localization threshold $\eta_c$. We can see that the point at $h=0$ is
absent, i.e. states are indeed delocalized. At the same time, the
function $f(h)$ has a continuous part, extending over the interval
$[h_\mathrm{min},1]$, while the lower edge $h_\mathrm{min}$ of the multifractal
spectrum is only slightly lower than $1$ for very weak disorder
$\eta=1$, but decreases fast when the disorder strength
increases. This can be interpreted as a presence of multifractal
extended states. We found it impossible do decide just on the basis
of available data, if it is only a finite-size effect or if it
survives in the limit $L\to\infty$.

In Fig.  \ref{mul-sp-loc} we show the
multifractal spectrum for strong disorder, close and above the  expected
localization threshold $\eta_c\gtrsim 6$. For the disorder $\eta=5$
the spectrum is continuous like in the Fig.  \ref{mul-sp-del}, but the
lower edge $h_\mathrm{min}$ is already very close to the point $h=0$
indicating localization. For $\eta=6$ and larger the point $h=0$ is
always present and when the disorder is increasing, the numerically
found points in the curve $f(h)$ more and more accumulate around $h=0$,
leaving only few points at $h>0$, which may ultimately be a single
point at $h=h_\mathrm{max}$, where the value is
$f(h_\mathrm{max})=1$.
This suggests the following
scenario in the limit $L\to\infty$. Exactly at the localization transition $\eta=\eta_c$ the
eigenvectors are multifractal, with $h_\mathrm{min}=0$. However, in
the localized phase,  $\eta>\eta_c$, the multifractal spectrum $f(h)$
is trivial and
collapses to just two points, $h=0$, $f(0)=0$, and $h=h_\mathrm{max}$,
$f(h_\mathrm{max})=1$.

\subsection{Spectral statistics}

\begin{figure}[t]
\includegraphics[scale=0.45]{%
\slaninafigdir/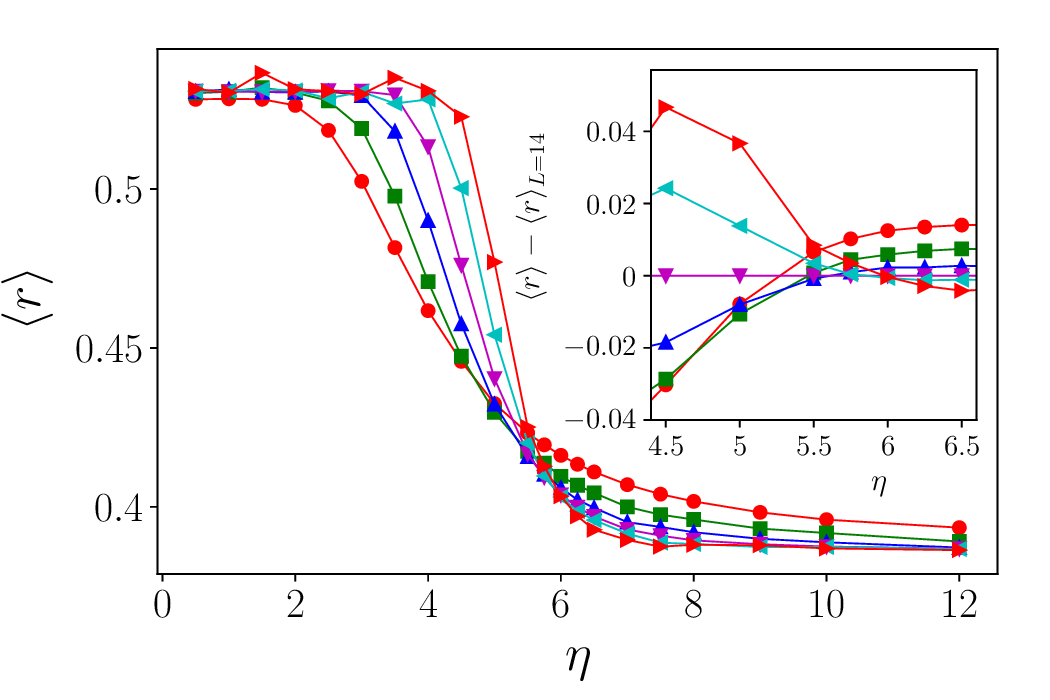}
\caption{Dependence of the averaged spacing ratio on the disorder
strength. The system size is $L=8$ ({\Large $\bullet$}), $10$ ($\blacksquare$), $12$ ($\blacktriangle$),
$14$ ($\blacktriangledown$), $16$ ($\blacktriangleleft$), $18$
($\blacktriangleright$). In the inset, we show the detail of the
area where the curves cross. For better visibility, we subtracted the
data for $L=14$.
}
\label{av-r}
\end{figure}

One of the widely used signatures of localization is the change in the
distribution of level spacings from Poisson on localized side to
Wigner-like on the extended side. The aggregate quantity which
discriminates between the two is the averaged spacing ratio
\begin{equation}
\langle r\rangle
=\Big\langle
\frac{
\min(\epsilon_{i+1}-\epsilon_i,\epsilon_i-\epsilon_{i-1})
}{\max(\epsilon_{i+1}-\epsilon_i,\epsilon_i-\epsilon_{i-1})
}
\Big\rangle
\end{equation}
where $\epsilon_i$ are the eigenvalues in ascending order and the
averaging is performed over the realizations of disorder and
random graph and also over narrow interval of energies around the
center of the band $\epsilon=0$. More precisely, for given $L$ we used
$3\cdot 2^{L-8}$ eigenvalues closest to $\epsilon=0$. We show the dependence of
$\langle r\rangle$ on disorder strength for several sizes
$L\le 18$ in Fig. \ref{av-r}. For determination of the critical
disorder we would like the
curves cross in a single point. However, this is not the case, as
observed notoriously in many numerical studies
\cite{oga_hus_07,pal_hus_10,lui_laf_ale_15}. The
crossing point shifts with increasing $L$ to larger and larger values,
thus giving no more than a lower estimate for the critical disorder,
in our case $\eta_c>6$. Note that this is perfectly consistent with
the lower estimate deduced from the study of IPR.

\section{Entanglement entropy}

Contrary to a generic random graph, the construction used here, namely
the hypercubic embedding, enables us to define in a natural way the
density matrix of a subsystem of the whole system. The eigenvectors of
the Hamiltonian can be indexed naturally in terms of the $L$ binary
variables $\{s_i\}_{i=1}^L\in\{-1,1\}^L$. Then, for
$n=1+\sum_{i=1}^L\,2^{i-2}(s_i+1)$, we have $\psi_n(\epsilon)$ the
$n$-th element of the eigenvector $|\psi(\epsilon)\rangle$ corresponding to eigenvalue
$\epsilon$. Let us now divide the system in two halves of equal
size. Let the subsystem $A$ consist of the first $L/2$ spins, the
subsystem $B$ of the remaining ones. We define the reduced density
matrix of the subsystem $A$ only by taking trace over the degrees of
freedom of the subsystem $B$. If we apply it on the (normalized) pure
state $|\psi(\epsilon)\rangle$ of
the  whole system, we have
\begin{equation}
\rho_{An n'}=
\sum_{n''=0}^{2^{L/2}-1}\,\psi_{n+2^{L/2}n''}(\epsilon)\,\psi_{n'+2^{L/2}n''}(\epsilon)
\end{equation}
the matrix elements of the reduced density matrix of the subsystem
$A$. Denote $\lambda_n$, $n=1,\ldots,2^{L/2}$ eigenvalues of the
matrix $\rho_A$. Then, the entanglement entropy is
\begin{equation}
S_e=-\sum_{n=1}^{2^{L/2}}\,\lambda_n\ln\lambda_n\;.
\end{equation}
We omit the reference to subsystem $A$ in the notation, as it will be
always implicitly assumed throughout. Besides the aggregate quantity
of entanglement entropy, we shall look also at more detailed
information contained in the entanglement spectrum, more precisely the
averaged density of eigenvalues
\begin{equation}
\mathcal{D}(\lambda) =
\Bigg\langle
2^{-L/2}\,\sum_{n=1}^{2^{L/2}}\,\delta(\lambda-\lambda_n)\Bigg\rangle
\label{eq:spectrum}
\end{equation}
where average is taken over realizations of the disorder and the
random graph. Of course, the knowledge of entanglement spectrum gives
us the average entanglement entropy
\begin{equation}
\langle S_e\rangle=-2^{L/2}\int
\mathcal{D}(\lambda)\,\lambda\,\ln\lambda\;d\lambda\;.
\label{eq:entropyfromspectrum}
\end{equation}

For a fully random Hamiltonian, it was conjectured by Page
\cite{page_93} that
if the system has Hilbert space of dimension $2^L$, then the
entanglement entropy of the subsystem which is exact half of it, is
given by the following formula
\begin{equation}
S_P=\sum_{k=2^{L/2}+1}^{2^L}\frac{1}{k}+2^{-1-L/2}-\frac{1}{2}
\stackrel{L\to\infty}{\simeq} \frac{L}{2}\ln 2 - \frac{1}{2}\;.
\label{eq:pageentropy}
\end{equation}
It is commonly called the Page entropy and we shall use it as a
reference value for our calculations. For the entanglement spectrum,
it was shown
\cite{yan_cha_ham_muc_15,ger_nan_reg_16,pie_par_mar_pas_sca_17} that
for $L\to\infty$ it approaches the Mar\v{c}enko-Pastur  distribution
\begin{equation}
2^{-L/2}\mathcal{D}(2^{-L/2}\,\lambda) \stackrel{L\to\infty}{\simeq}
\frac{1}{2\pi}\sqrt{\frac{4-\lambda}{\lambda}}\;.
\label{eq:marcenkopastur}
\end{equation}
It can be easily checked that inserting the  Mar\v{c}enko-Pastur
distribution (\ref{eq:marcenkopastur})
into the expression (\ref{eq:entropyfromspectrum}) we
obtain exactly
\begin{equation}
\langle S_e\rangle=\frac{L}{2}\ln 2 - \frac{1}{2}
\end{equation}
which is just the asymptotic expression for the Page entropy.
Therefore, the Page entropy is the benchmark value for the
entanglement entropy in the delocalized state. The fundamental feature
is the volume scaling $\langle S_e\rangle=O(L^1),\,L\to\infty$.
On the other hand, it
is not as straightforward to provide similar benchmark for the
localized regime. We only expect the surface scaling
$\langle S_e\rangle=O(L^0),\,L\to\infty$. It means that entanglement
entropy has a finite limit
$\lim_{L\to\infty}\langle
S_e\rangle=S_{e\infty}<\infty$ and we expect that the asymptotic
value as function of the disorder strength diverges at the
transition, i.e. $\lim_{\eta\to\eta_c+} 1/S_{e\infty}(\eta)=0$.

\begin{figure}[t]
\includegraphics[scale=0.45]{%
\slaninafigdir/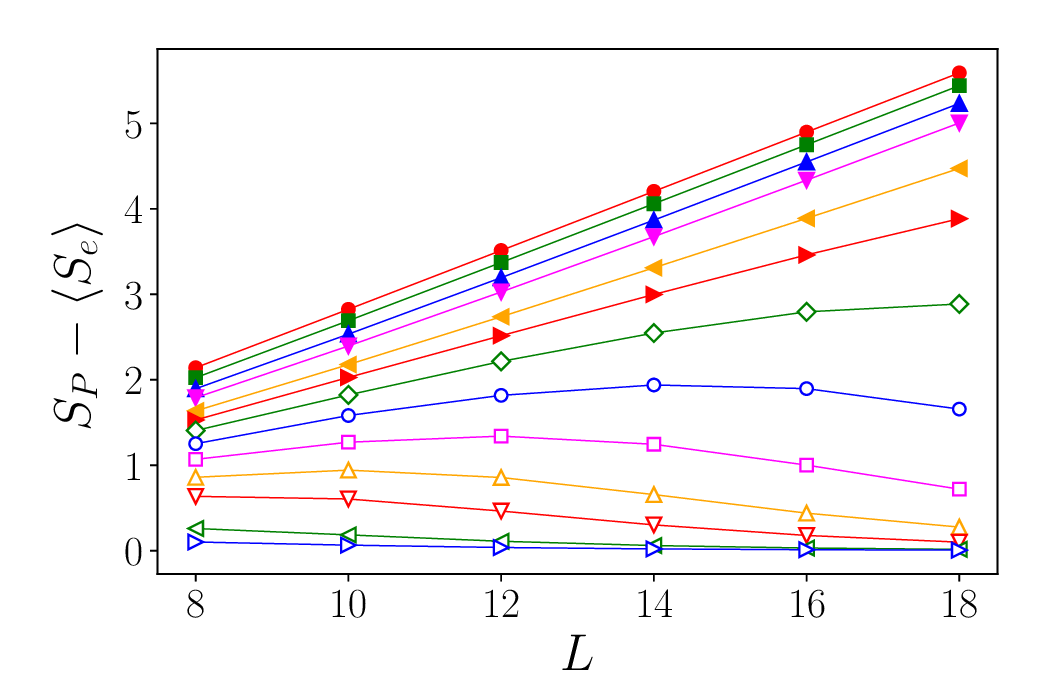}
\caption{Size dependence of the difference between Page entropy and
entanglement entropy for disorder strengths
$\eta=14  $ ({\Large $\bullet$}),
$10  $ ($\blacksquare$),
$8   $ ($\blacktriangle$),
$7   $ ($\blacktriangledown$),
$6   $ ($\blacktriangleleft$),
$5.5 $ ($\blacktriangleright$),
$5   $ ({\Large $\diamond$}),
$4.5 $ ({\Large $\circ$}),
$4   $ ($\Box$),
$3.5 $ ($\bigtriangleup$),
$3   $ ($\bigtriangledown$),
$2   $ ({\Large $\triangleleft$}),
$1   $ ({\Large $\triangleright$}).
}
\label{enta-entr}
\end{figure}

In Fig. \ref{enta-entr} we show the size dependence of the difference
of the entanglement entropy averaged over realizations, minus the Page
entropy (\ref{eq:pageentropy}). We can see that for weak disorder this
difference tends to zero, while for sufficiently large disorder it
increases linearly. This distinguishes the delocalized and localized
regime. However, we observe once again the same phenomenon in the
dependence on $L$, namely that for small $L$ the difference seems
increasing, thus giving false impression of being in the localized
phase, but then reaches a local maximum and eventually decreases
toward zero. Therefore, this diagram can only provide a lower bound on
the critical disorder strength, in this case it is even weaker than
before. From Fig. \ref{enta-entr} we can estimate that $\eta_c>5$.

\begin{figure}[t]
\includegraphics[scale=0.45]{%
\slaninafigdir/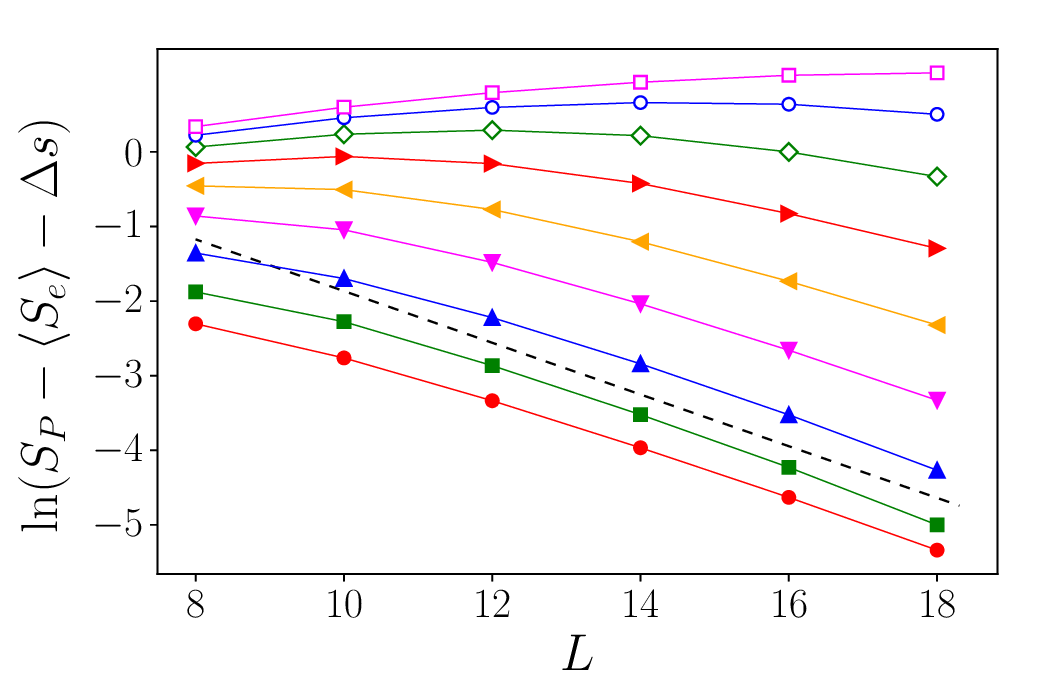}
\caption{Exponential approach to the Page entropy with increasing
system size, in the delocalized phase. The disorder strengths
are
$\eta=1  $ ({\Large $\bullet$}),
$1.5  $ ($\blacksquare$),
$2   $ ($\blacktriangle$),
$2.5   $ ($\blacktriangledown$),
$3   $ ($\blacktriangleleft$),
$3.5 $ ($\blacktriangleright$),
$4   $ ({\Large $\diamond$}),
$4.5 $ ({\Large $\circ$}),
$5   $ ($\Box$). The dashed line is the dependence
$S_p-S_e-\Delta  s\propto e^{-(\ln 2)L/2}$.
}
\label{enta-entr-del-phase}
\end{figure}

However, further information can be obtained from the approach to
asymptotic values for $L\to\infty$. The approach to Page entropy in
the delocalized phase is shown in Fig. \ref{enta-entr-del-phase}. We
found that
\begin{equation}
S_P-\langle S_e\rangle\simeq \Delta s+ c_\mathrm{del}(\eta)e^{-(\ln
2)L/2}\,,L\to\infty
\label{eq:appoachdel}
\end{equation}
where the constant shift $\Delta s=0.003$ was found empirically so that
the rest is asymptotically as close to exponential as possible.

\begin{figure}[t]
\includegraphics[scale=0.45]{%
\slaninafigdir/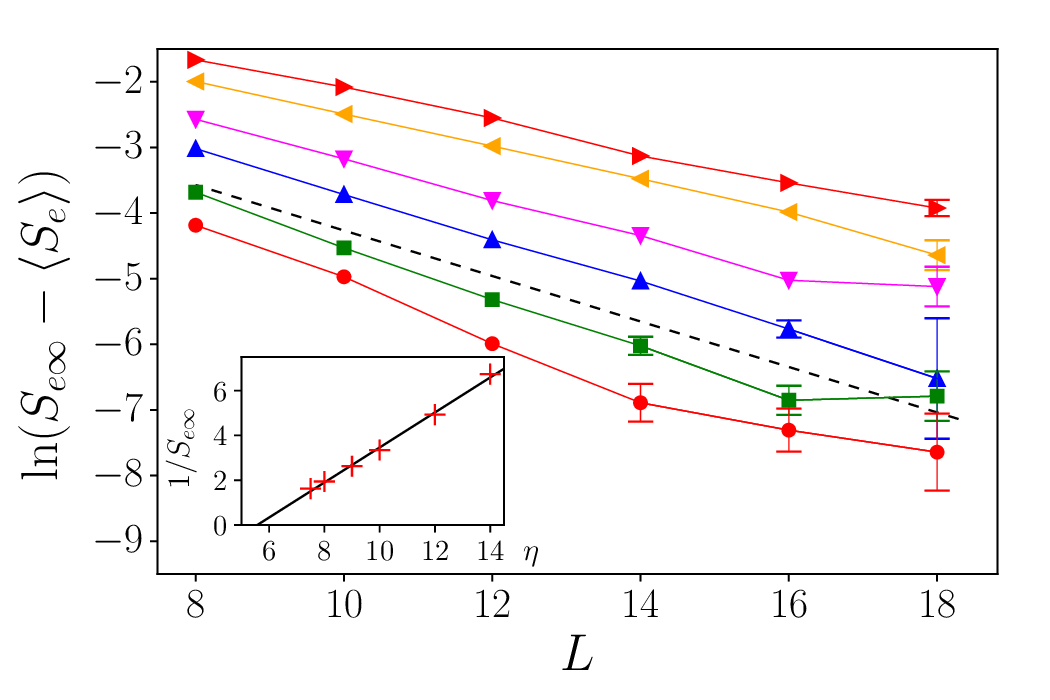}
\caption{Exponential approach of the entanglement entropy to the
asymptotic
constant $S_{e\infty}$  with increasing
system size, in the localized phase. The disorder strengths
are
$\eta=14  $ ({\Large $\bullet$}),
$12  $ ($\blacksquare$),
$10   $ ($\blacktriangle$),
$9   $ ($\blacktriangledown$),
$8   $ ($\blacktriangleleft$),
$7.5 $ ($\blacktriangleright$). Where not shown, the error bars are
smaller than the symbol size.
The dashed line is the dependence
$S_{e\infty}-S_e\propto e^{-(\ln  2)L/2}$.
In the inset, the dependence of the
inverse of the asymptotic value of the entanglement entropy on
disorder strength. The straight line fits the data and serves to
estimate the critical point $\eta_c$ of the localization transition.
}
\label{enta-entr-loc-phase}
\end{figure}

Similarly, on the localized side, we show in
Fig. \ref{enta-entr-loc-phase} the approach of the entanglement
entropy to the asymptotic constant value
\begin{equation}
S_{e\infty}(\eta)-\langle S_e\rangle\simeq
c_\mathrm{loc}(\eta)e^{-(\ln 2)L/2}\,,L\to\infty\;.
\label{eq:appoachloc}
\end{equation}
The
asymptotic constants $S_{e\infty}(\eta)$ provide an information
about the localization transition. As shown in the inset of
Fig. \ref{enta-entr-loc-phase}, the inverse  $1/S_{e\infty}(\eta)$
approaches linearly zero  at value $\eta_c=5.6(1)$. We can see that
this estimate gives a value which is slightly below the lower estimate
obtained e.g. from the study of IPR. The explanation might be that the
approach of $1/S_{e\infty}(\eta)$ is actually sublinear, rather than
linear in the vicinity of the transition.

\begin{figure}[t]
\includegraphics[scale=0.45]{%
\slaninafigdir/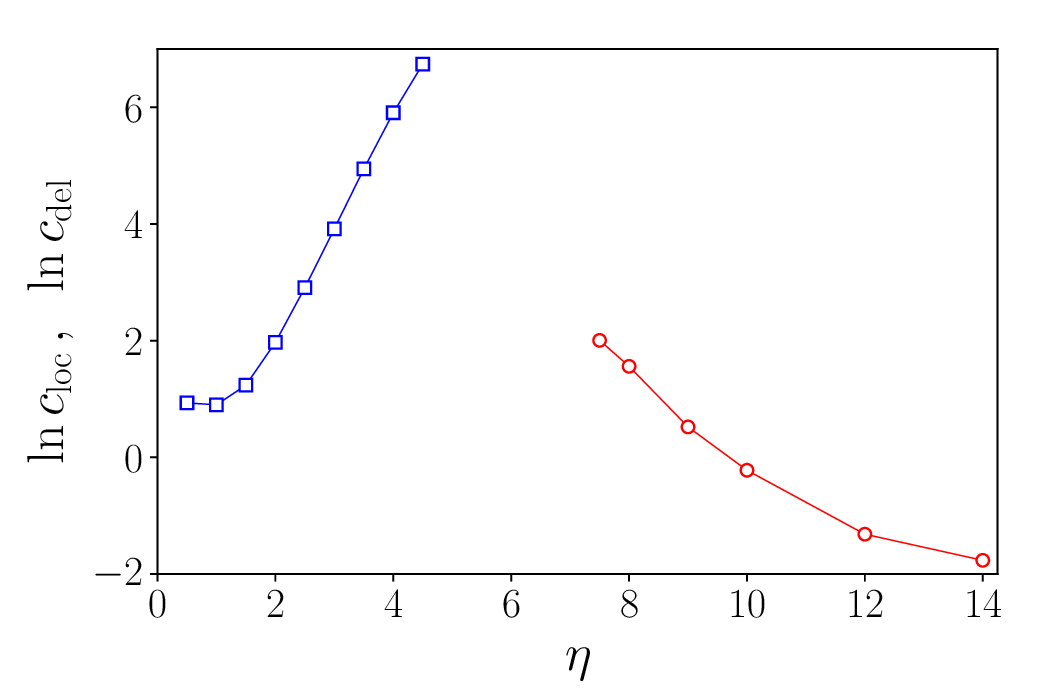}
\caption{Dependence of the coefficients in the finite-size corrections
(\ref{eq:appoachdel}) and (\ref{eq:appoachloc})
on the disorder strength. The symbols distinguish $c_\mathrm{del}$
($\Box$)) and  $c_\mathrm{loc}$ ({\Large$\circ$}).
}
\label{prefactors}
\end{figure}

We expect that the coefficients  $c_\mathrm{del}(\eta)$ and  $c_\mathrm{loc}(\eta)$
of the finite-size corrections diverge when we approach to the
transition. We show in Fig \ref{prefactors} their dependence on the
disordered strength. Indeed, we can see that both of these
coefficients increase in expected sense. On this basis, we can deduce
weak bounds on the critical disorder $5<\eta_c<7.5$,
but the assumed divergence is
not clear enough to provide better quantitative estimate of the value of
the critical disorder.

Interestingly, the approach follows the same exponential law as in the
delocalized phase. In fact, when we realize that $N_A=2^{L/2}$ is the
dimension of the Hilbert space of the subsystem $A$, we recognize in
both the exponential laws in (\ref{eq:appoachdel}) and
(\ref{eq:appoachloc}) the leading $1/N_A$ correction.


More detailed information is contained in the entanglement spectrum
(\ref{eq:spectrum}). It was already noted that the spectrum
qualitatively changes its character when we go from the delocalized to
localized phase
\cite{yan_cha_ham_muc_15,ser_mic_aba_pap_16,ger_nan_reg_16,ger_reg_nan_17,bui_gri_che_19,pie_par_mar_pas_sca_17}.
In the work \cite{pie_par_mar_pas_sca_17} the authors studied the
changes in the shape of the entanglement spectrum when the transition
is approached from the delocalized phase and used it to calculate the
correlation length. We shall use the methodology of
\cite{pie_par_mar_pas_sca_17} to see the qualitative change. In
delocalized phase, the spectrum $\mathcal{D}(\lambda)$
approaches the Mar\v{c}enko-Pastur
shape in the thermodynamic limit. It is characterized by
$\lambda^{-1/2}$ singularity for $\lambda\to 0$. The value of the
exponent is better extracted using the logarithmic derivative
\begin{equation}
\beta(\lambda)=\frac{\lambda}{\mathcal{D}}\,\frac{d\mathcal{D}(\lambda)}{d\lambda}\;.
\end{equation}
In delocalized phase, and for large $L$, the spectrum is supposed to
be close to Mar\v{c}enko-Pastur, which has
\begin{equation}
\beta_{MP}(\lambda)=-\frac{2}{4-\lambda 2^{L/2}}
\label{eq:betamp}
\end{equation}
and indeed the exponent is $\beta(\lambda)\to -1/2$ for
$\lambda\to 0$.

\begin{figure}[t]
\includegraphics[scale=0.45]{%
\slaninafigdir/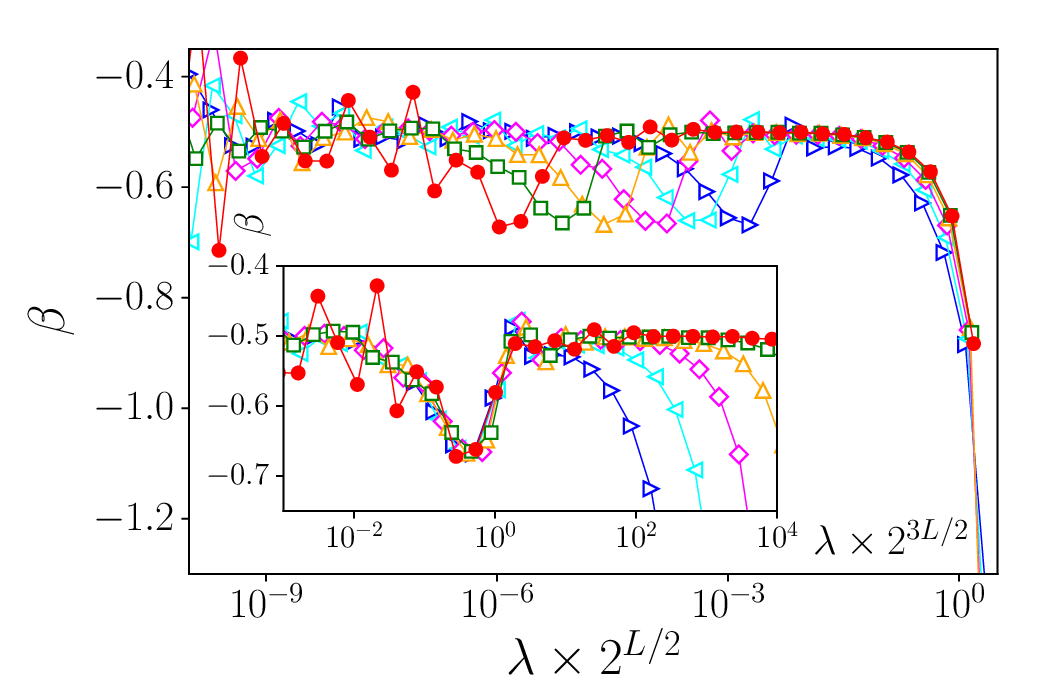}
\caption{Logarithmic derivative of the entanglement spectrum, for
$\eta=1$. In the inset, the same data scaled differently. The
symbols correspond to the sizes $L=8$ ({\large $\rhd$}), $ 10$ ({\large $\lhd$}), $12$
({\Large $\diamond$}), $14$ ($\triangle$), $16$ ($\Box$), $18$ ({\Large $\bullet$}).
}
\label{beta-dis1}
\end{figure}

We show in Fig. \ref{beta-dis1} the numerically obtained function
$\beta(\lambda)$ for weak disorder $\eta=1$ which is supposed to be
deep in the delocalized phase.
The formula (\ref{eq:betamp}) suggests the scaling
$\lambda\sim  2^{-L/2}$ of the characteristic eigenvalues, therefore
we plot $\beta$ as function of $\lambda 2^{L/2}$ for different sizes
$L$. We can see that indeed, with such rescaling of $\lambda$ the
spectrum approaches very well to the Mar\v{c}enko-Pastur
formula. However, we notice also a systematic contribution which
vanishes in thermodynamic limit when plotted in scaling
$\lambda 2^{L/2}$, but remains stable, when the function $\beta$ is
plotted against the variable $\lambda 2^{3L/2}$. It implies that
besides the universal characteristic scale $\lambda\sim  2^{-L/2}$
there is also another, model-dependent characteristic scale
$\lambda\sim  2^{-3L/2}$. This should reveal itself in the
next-to-leading finite-size corrections to the entanglement
entropy. However, this is out of reach of the current numerical
results. Note that similar feature observed in
Ref. \cite{pie_par_mar_pas_sca_17} had consequences already in the
leading finite-size corrections.

\begin{figure}[t]
\includegraphics[scale=0.45]{%
\slaninafigdir/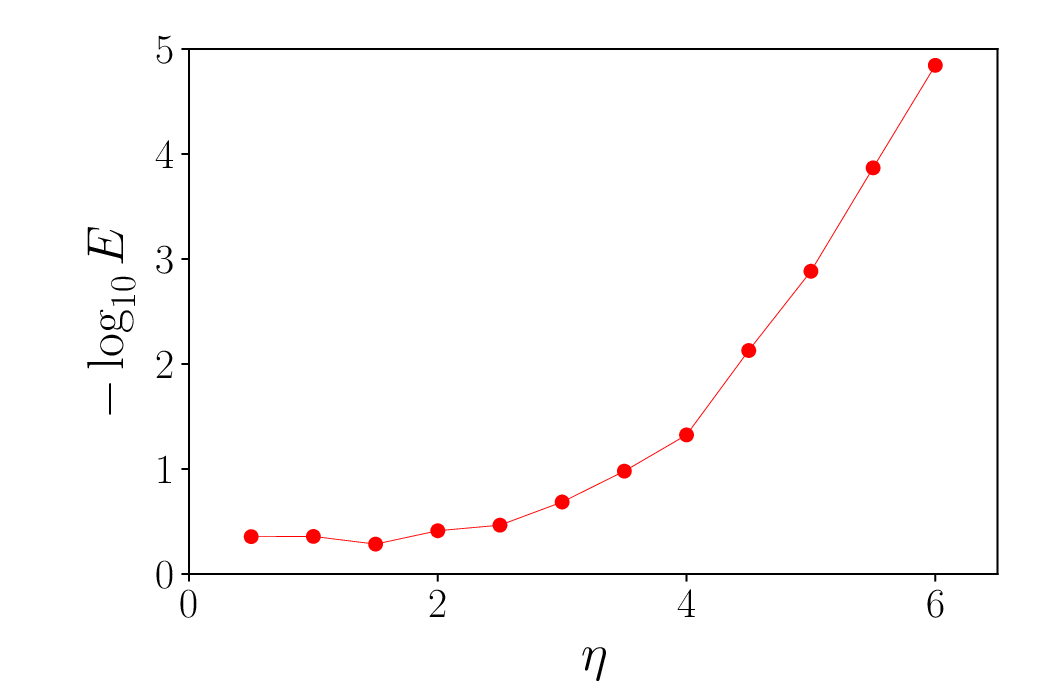}
\caption{ {Logarithm of the position of the minimum of
$\beta$ in the
entanglement spectrum, depending on the disorder strength. For
values $\eta>6$ there are no data, because the minimum disappears in
localized phase.}
}
\label{min-vs-eta}
\end{figure}

 {
We found that this characteristic energy scale reflects the approach
to  the localization transition. Indeed, we denote
$E=\lambda_\mathrm{min}\, 2^{3L/2}$ the position of the
minimum of $\beta$ seen in inset of Fig. \ref{beta-dis1}. Then, the
value of $-\log_{10}E$ seems to diverge when we approach to the
critical disorder, as shown in Fig. \ref{min-vs-eta}. In the localized phase the
minimum disappears completely. This suggests  that there is a
typical time scale $1/E$ that goes to infinity at localization threshold.
}

\begin{figure}[t]
\includegraphics[scale=0.45]{%
\slaninafigdir/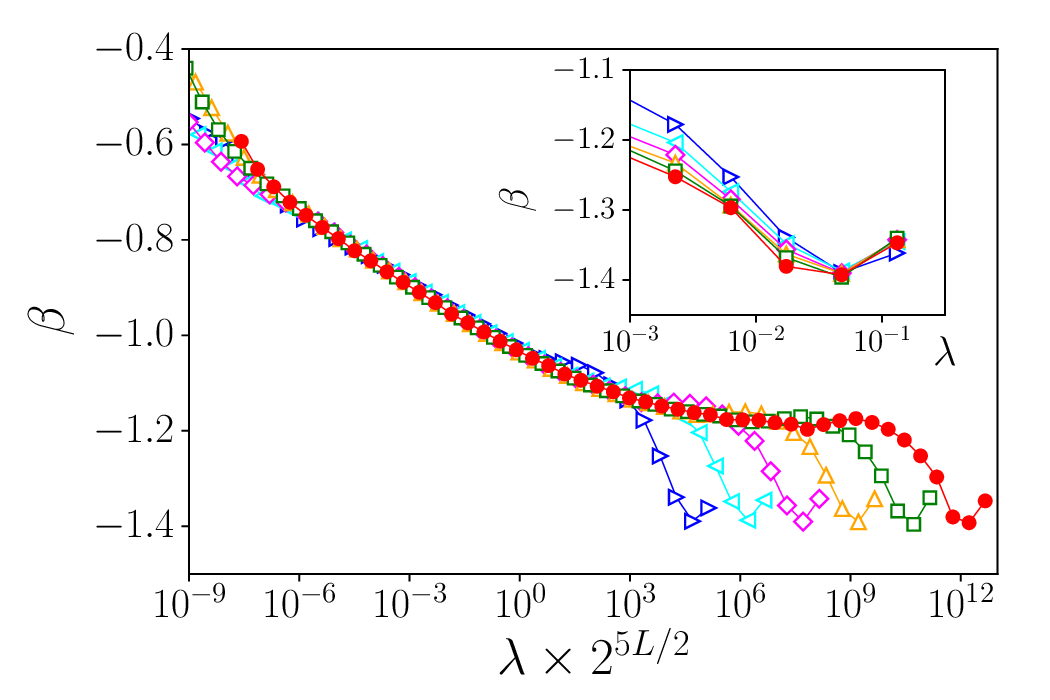}
\caption{Logarithmic derivative of the entanglement spectrum, for
$\eta=12$. In the inset, the same data scaled differently. The
symbols correspond to the sizes $L=8$ ({\large $\rhd$}), $ 10$ ({\large $\lhd$}), $12$
({\Large $\diamond$}), $14$ ($\triangle$), $16$ ($\Box$), $18$ ({\Large $\bullet$}).
}
\label{beta-dis12}
\end{figure}

In the localized phase, the spectrum looks completely different. We
show in Fig. \ref{beta-dis12} the numerically obtained function
$\beta(\lambda)$ for strong disorder $\eta=12$ which is supposed to be
deep in the localized phase. In absence of a clear guiding principle
as to the characteristic scaling of the eigenvalues, we proceeded
empirically and found that the best data collapse is achieved when
$\beta$ is plotted against $\lambda 2^{5L/2}$. The data collapse
stretches over many decades. However, at the upper edge of the
spectrum we found that $\beta$ is simply a function of $\lambda$, as
shown in the inset of Fig.  \ref{beta-dis12}. Having in mind that it
is the upper edge of the spectrum that dominates the entanglement
entropy (\ref{eq:entropyfromspectrum}), the observed scaling
$\lambda\sim (2^{-L/2})^0$ at the upper edge is consistent with surface
scaling $\langle S_e\rangle=O(1)$ for $L\to\infty$.

\section{Eigenstate thermalization hypothesis}

Localization can be viewed either as suppression of transport or as
broken ergodicity or as lack of thermalization. The latter approach is
formalized in terms of violation of the Eigenstate thermalization
hypothesis  (ETH)
\cite{deutsch_91,srednicki_94,rig_dun_ols_08,reimann_15}. Stated in
rather simplified formulation, ETH assumes that matrix elements of
local observables $\mathcal{O}$ in any narrow interval in highly excited part of
the spectrum behave in the following generic way
\begin{equation}
\begin{split}
\langle\psi_n|\mathcal{O}|\psi_m\rangle=
g(\epsilon_n)\delta_{nm}\\
+\xi_{mn}\,h(\frac{1}{2}(\epsilon_n+\epsilon_m),\epsilon_n-\epsilon_m)
\end{split}
\label{eq:ethformula}
\end{equation}
where $\xi_{mn}$ are normally distributed random variables and
$g(\epsilon)$ and $h(\epsilon,\omega)$ are smooth functions of their
arguments. It is an analog and generalization of the spectral
hypothesis stating that a portion of the spectrum in highly excited states
behaves like a spectrum of a random matrix and it depends only on the
symmetry of the system, whether it belongs to orthogonal, unitary or
symplectic ensemble, while the details of the Hamiltonian are
irrelevant. In the case of ETH, the basic notion behind is the
equivalence of statistical ensembles. In addition to canonical and
microcanonical ensembles, we consider also eigenstate ensemble, which
is nothing else than a pure state of the system. If we consider a
local observable, we can think of a part of the system containing the
support of that observable as a small but still macroscopic
subsystem. If the equivalence of ensembles holds for the whole
system, specifically if microcanonical and eigenstate ensembles were
equivalent, the average of the local observable would be the same in
both cases, with Gaussian fluctuations on top of it. This is in words
what (\ref{eq:ethformula}) expresses as formula.

ETH may fail in a number of ways. The most obvious is that the
functions $g(\epsilon)$ and $h(\epsilon,\omega)$ are not smooth, but
instead fluctuate wildly, giving in fact very different values for each
eigenstate. Another way of breaking ETH is having broad or long-tailed
distribution of the random variables $\xi_{mn}$, instead of a Gaussian one.
To see if ETH is broken or not we investigated the histograms of the
diagonal and off-diagonal elements of the spin
operators, defined in
(\ref{eq:localspins}).
Let us denote the matrix element of the $z$-component of the  spin operator at site $i$ as
$\sigma_{iznm}=\langle\psi_n|\sigma_{iz}|\psi_m\rangle$. We take the
eigenvectors corresponding to energies within a narrow interval
$I_{\Delta}=[-\Delta,\Delta]$ around the
center of the band. Actually the interval was fixed so that it contained
$N_\Delta=3\cdot 2^{L-8}$ eigenvalues closest to zero.
This distribution is subsequently averaged over site $i=1,\ldots,L$,
and finally also over realizations of the disorder. So, we define
\begin{equation}
P_\mathrm{diag}(\sigma)=\Bigg\langle\frac{1}{L}\sum_{i=1}^L\frac{1}{N_\Delta}\sum_{\substack{n
\\ \epsilon_n\in I_{\Delta}}}
\delta(\sigma- \sigma_{iznn})\Bigg\rangle
\end{equation}
for the diagonal elements and similarly
\begin{equation}
P_\mathrm{off}(\sigma)=\Bigg\langle\frac{1}{L}\sum_{i=1}^L\frac{1}{N_\Delta-1}\sum_{\substack{n
\\ \epsilon_n\in I_{\Delta}\\ \epsilon_{n+1}\in I_{\Delta}}}
\delta(\sigma- \sigma_{iznn+1})\Bigg\rangle
\end{equation}
for the off-diagonal matrix
elements between adjacent energy levels. Of course, both of these
distributions are defined on the interval $-1\le\sigma\le 1$.

\begin{figure}[t]
\includegraphics[scale=0.8]{%
\slaninafigdir/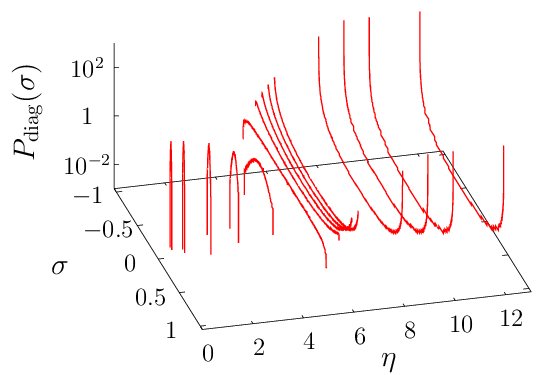}
\caption{Distribution of the values of the diagonal elements of the
local spin operators, for several disorder strengths $\eta$.
}
\label{hist-spin-diag-3d-d18-gp}
\end{figure}

\begin{figure}[t]
\includegraphics[scale=0.45]{%
\slaninafigdir/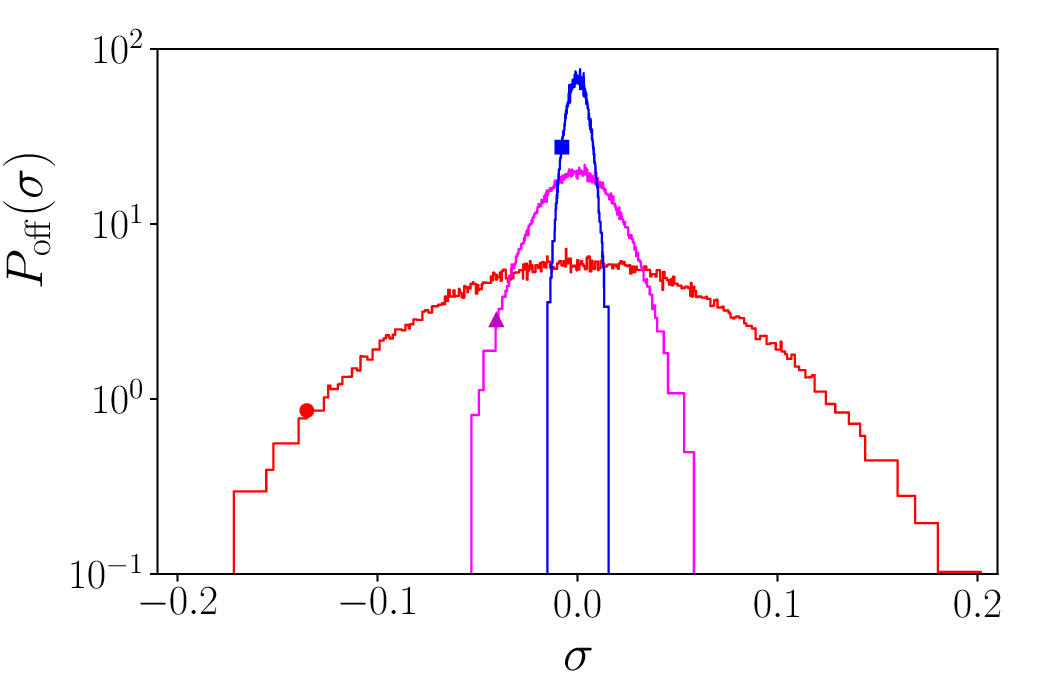}
\caption{Distribution of the values of the off-diagonal elements of the
local spin operators, for disorder strength $\eta=1$. Different
curves belong to different system sizes, according to the
attached symbols, $L=10$ ({\Large $\bullet$}), $14$
($\blacktriangle$), and $18$ ($\blacksquare$).
}
\label{hist-spin-offdiag-d10d14d18-eta1}
\end{figure}

\begin{figure}[t]
\includegraphics[scale=0.45]{%
\slaninafigdir/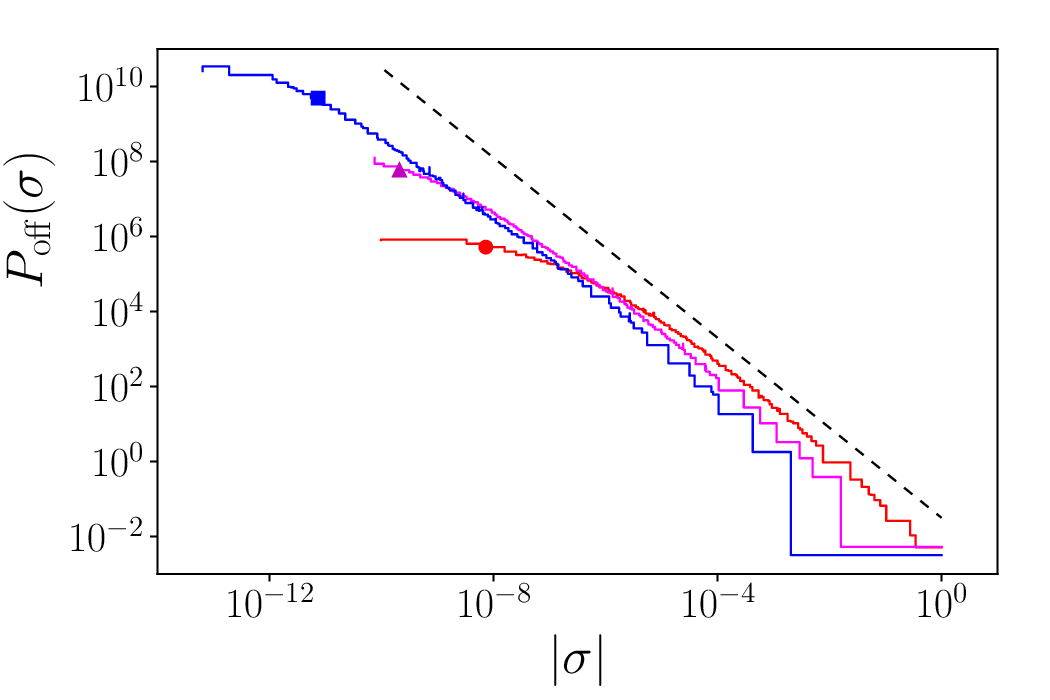}
\caption{Distribution of the values of the off-diagonal elements of the
local spin operators, for disorder strength $\eta=12$. Different
curves belong to different system sizes,  according to the
attached symbols,  $L=10$ ({\Large $\bullet$}), $14$
($\blacktriangle$), and $18$ ($\blacksquare$). The dashed line is
the power $\propto|\sigma|^{-1.2}$.
}
\label{hist-spin-offdiag-d10d14d18-eta12}
\end{figure}

We show in Fig. \ref{hist-spin-diag-3d-d18-gp} histograms of the
diagonal elements fro the largest size studied, $L=18$ and for
various disorder strengths. We can clearly see that the distribution
is centered at zero and close to Gaussian for weak disorder. Close to
the transition the distribution widens and starts to cover the whole
allowed interval $\sigma\in[-1,1]$. In the localized phase, the
distribution totally changes character and becomes bimodal, peaked
around the two edge values  $\sigma=-1$ and  $\sigma=1$. This is a
clear sign of violation of ETH. Note that similar behavior was already
observed in earlier works
\cite{pal_hus_10,luitz_16,pan_sca_sch_tay_zni_19,goi_eis_kru_19}.
We looked also on the off-diagonal elements. In
Fig. \ref{hist-spin-offdiag-d10d14d18-eta1} we show the histogram for
weak disorder, $\eta=1$, and three system sizes. We can see that the
distribution is close to Gaussian and moreover, its width decreases
with increasing size. The same size dependence was observed also in
the distribution of diagonal elements at weak disorder (not
shown). This indicates, that ETH as formulated by the ansatz
(\ref{eq:ethformula}) is satisfied in the delocalized phase and the
variance of the random variables $\xi_{nm}$ decreases in the
thermodynamic limit. In the localized phase, the off-diagonal elements
behave differently from the diagonal ones. In
Fig. \ref{hist-spin-offdiag-d10d14d18-eta12} we show the histogram for
$\eta=12$ and three different sizes. We can see, that the distribution
is again centered at zero as in the delocalized phase, but the shape
of the distribution is no more Gaussian, but close to a power law
$P_\mathrm{off}(\sigma)\propto |\sigma|^{-\gamma}$. For
the data in Fig. \ref{hist-spin-offdiag-d10d14d18-eta12} we estimated
the exponent about $\gamma\simeq 1.2$. However, this value seems to
depend on the disorder strength $\eta$, so it cannot be considered a
universal exponent.

The above observations suggest the following scenario of the violation
of the ETH as stated in (\ref{eq:ethformula}), when we go from the
delocalized into the localized phase. First,  in the
limit $L\to \infty$, the function
$g(\epsilon)$ is asymptotically close to $0$ for all energies in the
delocalized phase. On the other hand
$g(\epsilon)\in\{-1,1\}$ in the localized phase and which of the two
values holds for a given energy level $\epsilon_n$ is essentially a
random choice with equal probability. Moreover, the random variables
$\xi_{nm}$ are normally distributed, as assumed in ETH, in the
delocalized phase, but become heavy-tailed, probably power-law
distributed, in the localized phase.


\begin{figure}[t]
\includegraphics[scale=0.45]{%
\slaninafigdir/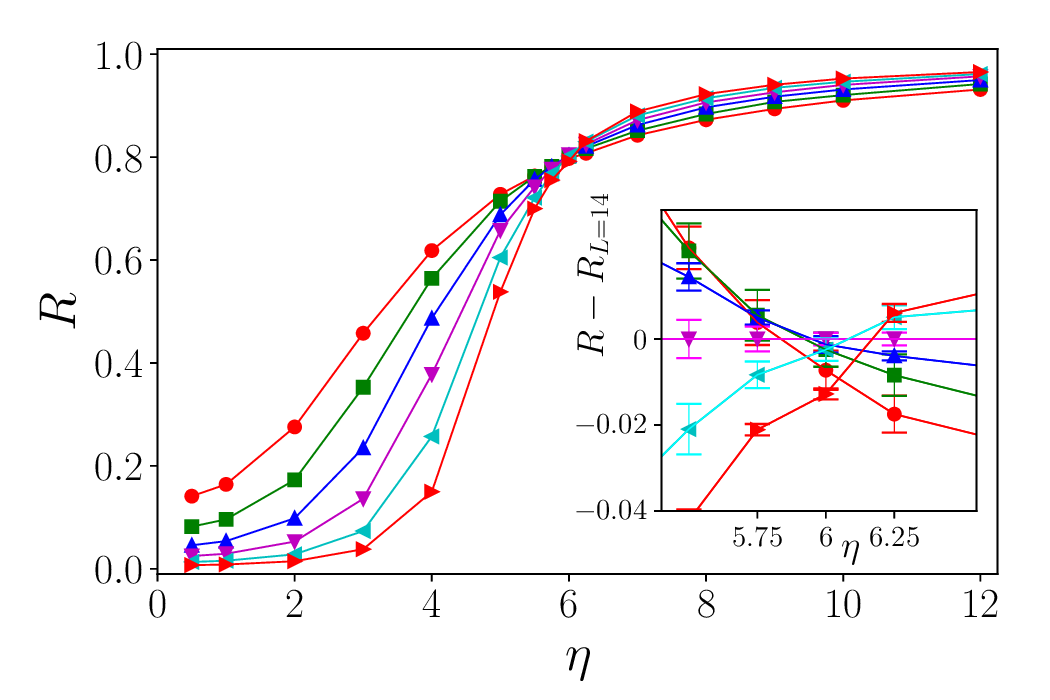}
\caption{Level-to-level fluctuations of the
$z$-component of the local spin operator, as a function of the
disorder strength. The levels were taken in a narrow interval
around the center of the band $\epsilon=0$.The system size is
$L=8$ ({\Large $\bullet$}), $10$ ($\blacksquare$), $12$
($\blacktriangle$),
$14$ ($\blacktriangledown$), $16$ ($\blacktriangleleft$), $18$
($\blacktriangleright$). In the inset, we show the detail of the
area where the curves cross. For better visibility, we subtracted the
data for $L=14$.
}
\label{eth}
\end{figure}

The distributions $P_\mathrm{diag}(\sigma)$ and
$P_\mathrm{off}(\sigma)$ are very instructive, but mix together the
level-to-level variability in matrix element (which is of main concern
here) with site-to-site and even worse, with sample-to-sample
variability. These variations are better separated in the aggregate
quantity
\begin{equation}
\begin{split}
R= \Bigg\langle \frac{1}{L}\sum_{i=1}^L \Big(\frac{1}{N_\Delta}
\sum_{\substack{n    \\ \epsilon_n\in I_{\Delta}}}
\sigma_{iznn}^2  -\\
\big(\frac{1}{N_\Delta}
\sum_{\substack{n   \\ \epsilon_n\in I_{\Delta}}}\sigma_{iznn}
\big)^2 \Big)
\Bigg\rangle
\end{split}
\label{eq-defR}
\end{equation}
which first calculates variance of level-to-level fluctuations and
then makes site and disorder average. We show in Fig. \ref{eth} how it
depends on the disorder strength. In the delocalized phase the
level-to-level fluctuations are low, as the ETH predicts  smooth
dependence on energy in function $g(\epsilon)$, which for narrow
energy interval which we use means effectively independence on energy
level. The remaining variance is due to superimposed Gaussian
fluctuations, but we have already seen, that their variance diminishes
when $L\to\infty$. So, we expect low, asymptotically zero, value of
$R$ in the delocalized phase. This is indeed observed in
Fig. \ref{eth}. On the contrary, in the localized  phase, the values
are either $+1$ or $-1$ randomly scattered among the levels, so that
the level-to-level fluctuations are close to $1$. Hence, in
thermodynamic limit we expect $R(\eta)=1$ for $\eta>\eta_c$ and
$R(\eta)=0$ for $\eta<\eta_c$. The data shown in Fig. \ref{eth}
support this expectation. The picture resembles that of the level
spacings shown in Fig. \ref{av-r}, but we believe that $R$ is
principally better signature of the localization transition than the
spacing ratio $\langle r\rangle$.
First, $R$ is in principle directly measurable quantity, limited only
by the experimental accessibility of creating pure states.
From the side of numerical simulations, as is the case of this work,
$R$ is also a much better measure.
The curves of $R(\eta)$ for increasing size $L$ do not
cross in exactly the same point, but the shift of the crossing point
is substantially weaker than  it is in the case of $\langle r\rangle$,
as can be seen in comparison of insets in Figs  \ref{av-r} and
\ref{eth}. On the basis of Fig. \ref{eth} we estimate the critical
disorder strength as $\eta_c=6.1\pm 0.3$. Note that this value tightly
satisfies the estimated lower bounds which were established by
different procedures, namely by studying the IPR, the level spacing
and the entanglement entropy. Although neither the study of the
breakdown of ETH provides precise answers, we consider it the most
reliable approach of those used in this work.

\section{Conclusions}

We introduced and studied a model which simulates a disordered system of
interacting particles with spin $1/2$. The Hilbert space is
isomorphic to the set of vertices of a hypercube. The Hamiltonian of
this system is the sum of two parts. The first part is the adjacency matrix
of a random graph, which is a subset of the edges of a hypercube and
simultaneously it is a random regular graph with degree three, i.e. a
random cubic graph. The second part is diagonal with independent Gaussian random
numbers with tunable variance. Therefore, it can be seen as a
special kind of a random-graph analog of the Rosenzweig-Porter random
matrix ensemble.
{The comparison with the Rosenzweig-Porter (RP) ensemble
must be taken with great care, though. First, our model corresponds
rather to log-normal RP \cite{kha_kra_alt_iof_20}
than the classical RP ensemble. Second,
having most matrix elements strictly zero, as is our case, could
make a big
difference from having random non-negative numbers, as is the case
of log-normal RP.  And third, the transition in RP ensembles is governed
rather by the exponent in the scaling of matrix elements with size
\cite{kra_kha_cue_ami_15},
while here the governing parameter is the amplitude of diagonal
elements. This makes our model closer to the original work of
Rosenzweig and Porter \cite{ros_por_60} than to the more recent investigations
\cite{kra_kha_cue_ami_15,ven_cug_sch_tar_23}. Further comparison to
RP ensembles would
require deeper study which was not our aim here.}

The random graph in question is created starting from
a manifestly cubic graph by a sequence
of random rewiring moves that preserve the degree of all affected
vertices. This algorithm surely does not cover uniformly the whole set of random cubic
graphs embedded in hypercube, but we believe it samples the set
representatively enough.
{We tested that on a small system of dimension $L=5$
and we found no deviation from uniformly sampled set.}

The properties of the model were investigated by exact diagonalization
of many instances of the random Hamiltonian, for system sizes
$L\le 18$.  The main question was the
presence and properties of the localization transition. In this work
we view th localization transition from a purely static perspective as
an eigenstate transition, i.e. sudden qualitative change in the nature
of the eigenvectors, when the strength of the disorder increases. This
is reflected first in the inverse participation ratio (IPR). This way we
found the estimated phase diagram in the plane energy-disorder and
observed the mobility edge, which separates the localized and
delocalized regions. The disorder strength at which localization
appears is highest at the center of the energy spectrum, thus marking
the critical disorder $\eta_c$, beyond which all eigenvectors are
localized. From the study of IPR, we obtained the lower bound
$\eta_c>6$. Close to the supposed transition we found clear sign of
multifractality in eigenvectors. At the same time we consider the seemingly
extended interval of disorder strengths which support multifractality
in delocalized phase to be merely a finite-size effect and we expect
this interval shrinks when size is further increased. However, much
better numerical data would be necessary to see it with certainty.

We also used the classical signature of the change in the character of
eigenvectors as reflected  in the
organization of energy levels. Indeed, the side-effect of localization
is lifting the level repulsion which results in Poisson
level-spacing distribution rather than the Wigner-like which is
universal characteristic of delocalized phase. We observed the
transition with all the well-known drawbacks, namely the marked shift
of the predicted transition point to larger values when the system
size increases. From this point of view our model behaves identically
to both the random regular graphs and true interacting-particle models,
like the disordered Heisenberg chain.

{So, to this point our model reproduces what is known
from the study of random regular graphs. The main point of our work,
however, lies in extensions which are not accessible in random
regular graphs with no extra properties. Specifically, our graphs
allow, first,  definition of entanglement entropy, and, second,
study of the violation of eigenvalue thermalization hypothesis. The
study of these two properties is the core of our contribution. Let
us discuss them now.}

The change in the nature of eigenvectors is visible in the
entanglement entropy and entanglement spectrum. This appears when we
look at a subsystem of the whole system, typically at an exact half.
As the random graph we
use is embedded in a $L$-dimensional hypercube, it is straightforward
how to define the density matrix of the subsystem whose Hilbert space
has dimension $L/2$. Note that this would be hardly possible for a
generic random regular graph. We stress here the advantage of our
approach. The entanglement entropy shows clear volume scaling with
increasing $L$ when in the delocalized phase, while the localized
phase is characterized by surface scaling. Analyzing the flow diagram
for increasing $L$ in the delocalized phase we get a weaker lower
bound for critical disorder,
$\eta_c>5$. In the localized phase, we observed the divergence of the
asymptotic value of entanglement entropy in the form
$(\eta-\eta_c)^{-1}$ i.e. with the critical exponent $\nu=1$.
The critical disorder estimated this way
was $\eta_c=5.6\pm 0.2$, which is slightly lower that the lower bound
obtained from IPR. The possible explanation might be that the
divergence is actually weaker, governed by lower exponent $\nu<1$.

We are able to see not only the asymptotic
behavior for $L\to\infty$, but also the leading finite size
corrections which are exponential in $L$ and proportional to $2^{-L/2}$
in both the localized and delocalized phases. The factors of
proportionality are expected to diverge when we approach the critical
point from either of the two sides. We indeed observed that the
factors increase in both phases when coming closer to the transition, but
without  clear sign of the expected divergence. Larger systems would
be necessary to see that effect, but with current algorithms there is
little hope
that we could go significantly beyond the maximum used here, which was
$L=18$. Thus, we found that the behavior of finite size  corrections
is a good sign of the existence of two distinct phases, the localized and the
delocalized one, but it is not an efficient way to establish the
transition point. Similarly, we found that the form of the
entanglement spectrum clearly distinguishes between delocalized and
localized phase. The delocalized phase is characterized by the
Mar\v{c}enko-Pastur exponent $-1/2$ up to the cutoff which scales with
the system size as $O(2^{-L/2})$. In the localized phase it is
not possible to specify a single exponent, but rather the locally defined
exponent of the entanglement spectrum decreases continuously from
$-1/2$ to about $-1.2$ when the eigenvalues increase over several
decades up to the maximum value, which is of order $O(2^{-1})$
independently of system size. Unfortunately, the entanglement
spectrum  does not provide a quantitative tool for extracting
the critical disorder value.

In this respect, we found that the much more efficient way is the study of the
violation of the Eigenstate thermalization hypothesis (ETH). We found that the
delocalized and localized phases are characterized by substantially
different distribution of diagonal, as well as off-diagonal elements of
local spin operators. We found that both diagonal and off-diagonal
elements are normally distributed in the delocalized phase, in
agreement with ETH. Moreover, the width of the Gaussian shrinks with
increasing system size. On the contrary, in delocalized phase the
distribution of diagonal elements is bimodal and the off-diagonal
elements follow a power-law distribution. Therefore, we observe clear
qualitative signs of the breakdown of ETH in the localized
phase. Quantitatively, it is observed in the level-to-level
fluctuations of the diagonal elements. This quantity is negligible in
the delocalized phase but close to one in the delocalized
phase. Indeed, high level-to-level fluctuations imply failure of ETH
and lack of thermalization. Therefore, the quantity of these
fluctuations may serve as order parameter for the localization
transition. Contrary to the level-spacing statistics, here we do not
observe significant shift of the crossing value with increasing
size. Therefore, we consider the level-to-level fluctuations the best
signature of localizations of those studied in this work. On this
basis, we estimate the critical disorder as $\eta_c=6.1\pm 0.3$. This
value is tightly above the lower bounds established from the study of
IPR, so we consider these findings consistent.

One of the main motivations for this work came from the field of
Many-body localization (MBL). Summarizing our results, we conclude that we
observed many of the characteristic features of MBL in our model,
despite the fact that the random Hamiltonian of our spin model does
not follow from an explicit particle-particle interaction, but rather it is
taken from a very specific random matrix ensemble. This could imply
that MBL is a generic phenomenon in certain class of random
Hamiltonians. One of the keywords in the study of MBL is locality, or,
equivalently, partitioning to subsystems.
Indeed, ETH is formulated as a property of local operators, or
operators whose support is limited to a small subsystem of the
whole. Analogically, the entanglement entropy (EE) belongs to specific
subsystems. The fact, that this subsystem is usually taken as
exact half of the whole is mainly due to computational convenience and
at the same time to practical impossibility to work with really big
systems. Indeed, in abstract considerations the subsystem whose EE we
calculate should be much smaller than the whole system, but still
macroscopic. In practice, though, this is impossible to satisfy with systems
as small as $L\le 18$. However, the requirement of locality or
partitioning imposes only quite general limits on the random matrix
ensemble in question. The Hilbert space must be a tensor product of
certain number of parts, each belonging to one possible partition, or
to a support of one of the local operators. The Hamiltonian, then, is
random, but must conform with this tensor product. The model we
presented and studied in this work, is one simple example of such type
of random Hamiltonians. Certainly more elaborate and even more general
models may follow. We can for example think of two random graphs,
corresponding to partitions $A$ and $B$, and define a new graph simply
taking tensor product of these two graphs, with possible cutting of
certain percentage of links, in order to keep the average degree
properly scaled. Studying entanglement entropy in such system would be
in principle straightforward. Or, having system $A$ much smaller,
perhaps corresponding to one particle, or spin, only, this scheme
would enable to study ETH with local operators acting only on the
single particle in $A$.

{This idea leads to the system of two parallel random
regular graphs diagonally coupled to each other. As the random
regular graphs are accessible to analytic means, like the replica
trick, there is a hope that also such double random regular graph
might be solvable. So, the model studied by us here can provide a
starting point for deeper analysis.}  {We show in the
Appendix basic steps in the analytical study which may follow.}
We leave the investigation of such host of
questions to future research.

\begin{acknowledgments}
Computational resources were provided by the e-INFRA CZ project
(ID:90254), supported by the Ministry of Education, Youth and Sports
of the Czech Republic.
\end{acknowledgments}
\appendix
\section{Comparison of algorithms for graph creation}

\begin{figure}[t]
\includegraphics[scale=0.45]{%
\slaninafigdir/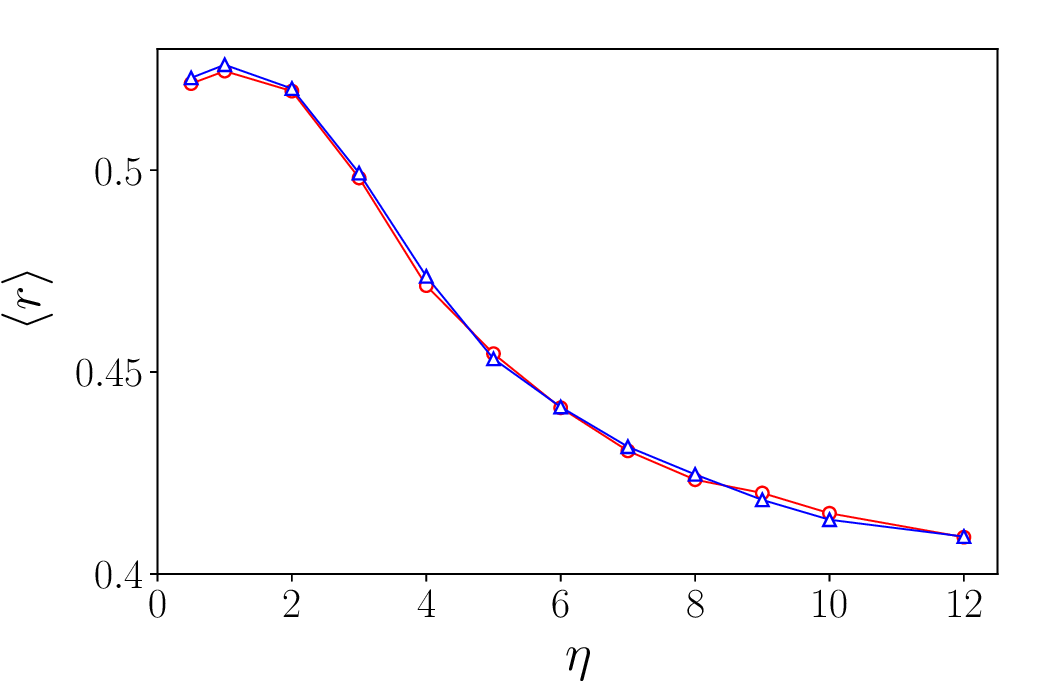}
\caption{{Dependence of the averaged spacing ratio
on the disorder
strength, for dimension $L=5$. The symbols {\Large$\circ$}
correspond to graphs created by rewiring algorithm, the symbols
$\triangle$ to graphs sampled uniformly.}
}
\label{av-r-L5}
\end{figure}

\begin{figure}[t]
\includegraphics[scale=0.45]{%
\slaninafigdir/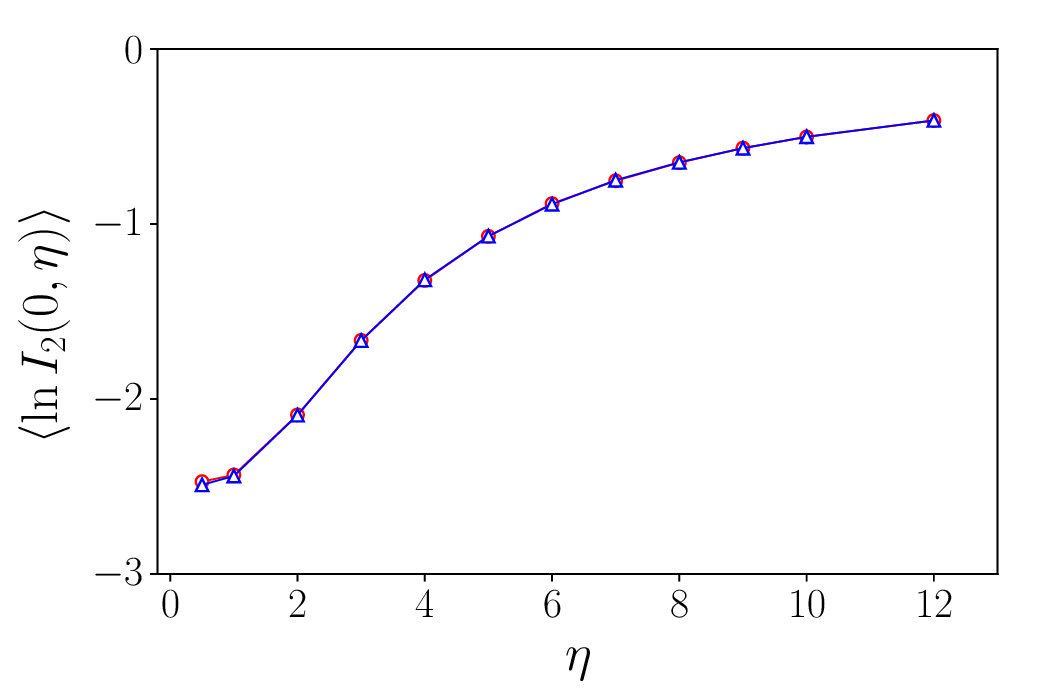}
\caption{{Dependence of the inverse participation
ratio at the center
of the spectrum on the
disorder strength, for dimension $L=5$. The symbols {\Large$\circ$}
correspond to graphs created by rewiring algorithm, the symbols
$\triangle$ to graphs sampled uniformly.}
}
\label{ipr-stred-L5}
\end{figure}

{
The number $H_3(L)$ of cubic graphs embedded into hypercube grows very fast
with dimension $L$. A very naive guess estimates  this
number  as $\sim L^{2^L}$. Without loss of generality we can fix that
the first vertex is
connected to three vertices belonging to first three of $L$ possible
directions. The counts of possibilities mentioned later are given with
such constraint.
Trivially $H_3(3)=1$ and it is not
difficult to find all such graphs for $L=4$.
We find $H_3(4)=68$. All
but one of these graphs are connected. Going one step further in
dimension is already computationally intensive task. For $L=5$ we were able to
enumerate all $H_3(5)=290701974$ cubic graphs embedded in
five-dimensional hypercube and we found that only very small fraction
of them are disconnected. The number of connected graphs was
$H^c_3(5)=290475585$. Proceeding to $L=6$ proved unfeasible (unless a
significantly smarter algorithm is available).
}

{
The graphs created by random rewiring algorithm described in the main
text do not sample the whole set of $H^c_3(5)$ graphs, but we
performed at least a partial check whether the rewiring-created graphs
are representative enough. To this end, we took the whole set of
$H^c_3(5)$ graphs we explicitely found, and we selected from it
samples randomly. This way we get uniformly-sampled set of graphs. We
compare its properties to rewiring-created set of graphs with $L=5$ by adding
random normally-distributed diagonal disorder to both sets and
calculating two basic properties of the spectrum, namely the spacing
distribution parameter $\langle r\rangle$ and the IPR in the middle of
the band (i.e. close to the eigenvalue $0$). We show in
Figs. \ref{av-r-L5} and \ref{ipr-stred-L5} how these quantities depend
on the disorder strength. We can see that the results for the
uniformly-sampled set and for the  rewiring-created set are nearly
indistinguishable one from the other. We consider that a strong
indication that the rewiring algorithm does not introduce any artifacts
which would impact on the results of our work.
}

 {
\section{Perspective of analytic calculations in dimer graph}
}

 {
Any graph embedded in a hypercube of dimension $L+1$ has natural dimer
structure. Indeed, its vertex set is a union of vertex sets of
two hypercubes of dimension $L$, while its edge set is the union of
three components. The first two components are the subsets of edges of
the first
and second $L$-dimensional hypercube, and the third component is the
subset of ``diagonal'' edges, which are those edges that connect
mutually homomorphic vertices in the first and the second
hypercube. This allows natural definition of the $z$-component of spin
operator as diagonal matrix with $+1$ on all vertices of the first
hypercube and $-1$ on all vertices of the second hypercube.
}

 {
We can generalize this idea to wider classes of graphs, beyond the
hypercube. Imagine two graphs homomorphic to each other in statistical
sense. Obviously, the two graphs must have the same number of
vertices. One example, which brings us back to the main subject of this
paper  are hypercubes with equal dimension and edges present or absent
with the same probabilistic manner in both. Another, more amorphous
example are two Erd\"os-R\'enyi graphs with the same edge
probability. Then, we can make a dimer from this pair of graphs by
joining the two by edges which are chosen from the set of ``diagonal''
edges. The existence of the homomorphism ensures the existence of such
set. Typically, if there is a homomorphism, there are more of
them. So, there is a multitude of ways to construct the dimer from two
homomorphic graphs. In any of these cases, it is possible to define
the $z$-component of spin
operator as explained above. This opens the road to studying
eigenvector thermalization hypothesis in random Hamiltonians based on
such dimer graphs.
}

 {
The appeal of such approach is in its potential analytic solvability.
We have no ambition to solve these problems in this paper, as our main
focus was only on a numerical study of a single specific example of
graphs which naturally have such dimer structure. However, we provide
here at least a sketch of possible pathways in further simplified
situations, where the Hamiltonian stemming from random graphs is
replaced by random matrix and the randomly placed zeros and ones
corresponding to ``diagonal'' edges are replaced by a constant number
$d$, which reflects the relative occurrence of ones. Moreover, we
limit ourselves only on extreme cases of fully localized and fully
delocalized cases. The fully localized case is straightforward and
provides an information on size dependence, while the fully
delocalized case needs the use of replica trick and provides just
infinite-size result.
}

 {
\subsection{Fully localized case}
}

 {
Let us denote the two components of the dimer $A$ and $B$. Suppose
that each of the components has $N$ vertices. The capital
indexes $I,J,\ldots\in\{A,B\}$ will denote components, lowercase indices
$i,j,\ldots\in\{1,2,\ldots,N\}$ will denote vertices within the components.
In fully localized phase the Hamiltonian can be approximated as
\begin{equation}
H_{IiJj}=\eta\,\delta_{IJ}\delta_{ij}\,\xi_{Ii}+d\,(1-\delta_{IJ})\delta_{ij}
\end{equation}
where  $\xi_{Ii}$ are
independent Gaussian random variables with unit variance. The
parameter $\eta$ quantifies the strength of the disorder. In the same
basis, the $z$-component of the spin operator is the matrix
\begin{equation}
\sigma_{zIiJj}=\delta_{IJ}\delta_{ij}(-1)^I
\end{equation}
with convention $(-1)^A=1$, $(-1)^B=-1$.
}

 {
Diagonalization of the Hamiltonian is trivial, as it splits into $N$
uncoupled $2\times 2$ matrices. Equally trivial is the calculation of
the average of the spin $\langle\sigma_z\rangle_\lambda$ in an
eigenstate with eigenvalue $\lambda$. The probability of this average
with respect of the Gaussian disorder $\xi_{Ii}$ and over all eigenvectors is
then
\begin{equation}
P(\langle\sigma_z\rangle)=
\frac{d}{\sqrt{\pi}\eta}
\frac{
\exp\Big(-\big(\frac{d}{\eta}\big)^2
\frac{\langle\sigma_z\rangle^2}{1-\langle\sigma_z\rangle^2}\Big)
}
{(1-\langle\sigma_z\rangle^2)^{3/2}}\;\;.
\end{equation}
}

 {
The mean of $\langle\sigma_z\rangle$ is of course zero due to
symmetry. The second moment can be computed in terms of the series in
powers of $(d/\eta)^2$, namely
\begin{equation}
\overline{\langle\sigma_z\rangle^2}=1-\frac{1}{2}
\Big(\frac{d}{\eta}\Big)^2+\frac{3}{16}\Big(\frac{d}{\eta}\Big)^4-\cdots\;.
\label{eq-avsigmasq-fulloc}
\end{equation}
}

 {
The dependence on the system size enters through the parameter $d$,
which measures the fraction of edges among all potential diagonal
links. In random cubic graph in a hypercube we have $N=2^L$, each
vertex has $L$ potential links within its own component and one diagonal
link. The diagonal link has therefore (for large $L$) probability
$\simeq 3/(L+1)$ to
be occupied by edge. This suggests the scaling $d\propto 1/\ln
N$. This should be compared with the behavior of the numerically
established quantity $R$ defined in (\ref{eq-defR}). On the basis of
(\ref{eq-avsigmasq-fulloc}) we expect that
\begin{equation}
1-R\sim \frac{1}{L^2\eta^2}
\end{equation}
for large $L$ and deep in the localized phase.
}

 {
\subsection{Sketch of the replica calculations in fully delocalized case}
}

 {
In contrast to the localized case, in the fully delocalized phase the
Hamiltonian can be approximated by a composition from two GOE random
matrices, connected diagonally as in the localized case. Therefore, we take
\begin{equation}
H_{IiJj}=\eta\,\delta_{IJ}\,A^{(I)}_{ij}+d\,(1-\delta_{IJ})\delta_{ij}
\end{equation}
where $A^{(I)}$ are $N\times N$ GOE random matrices with unit variance, i.e.
\begin{equation}
P(A^{(I)}_{ij})=
\sqrt{\frac{N}{2g\pi}}e^{-\frac{1}{2g}N(A^{(I)}_{ij})^2}
\end{equation}
and $g=1$ for $i\ne j$, $g=2$ for $i=j$.
}

 {
We can proceed in a standard manner, applying the replica method.
(From the numerous literature, let us mention
e.g. \cite{kam_mez_99,ven_cug_sch_tar_23} and references therein.)
The first step is the calculation of the resolvent $R(\zeta)=(\zeta-H)^{-1}$,
$\zeta\in\mathbb{C}$, which is directly related to the averaged density of
eigenvalues
\begin{equation}
\overline{\mathcal{D}}(E)=\frac{1}{\pi}\lim_{\epsilon\to 0^+}\mathrm{Im}
\overline{\mathrm{Tr}\,R(E-\mathrm{i}\epsilon)}
\end{equation}
(here and in the following the bar denotes average over disorder). We can
similarly calculate the quantity
\begin{equation}
\overline{\mathcal{D}_z}(E)=
\frac{1}{\pi}\lim_{\epsilon\to 0^+}\mathrm{Im}
\overline{\mathrm{Tr}\,\sigma_zR(E-\mathrm{i}\epsilon)}
\end{equation}
and from it deduce the average of the $z$-component of the spin
operator at energy $E$
\begin{equation}
\overline{\langle\sigma_z\rangle}_E=
\frac{\overline{\mathcal{D}_z}(E)}{\overline{\mathcal{D}}(E)}\;.
\end{equation}
Note the seemingly inconsistent replacement of average of a fraction
by fraction of averages. In fact, for however large but finite $N$
this is the only
sensible way to compute $\overline{\langle\sigma_z\rangle}_E$, because
before averaging neither $\mathcal{D}_z(E)$ nor $\mathcal{D}_z(E)$ are
continuous functions of $E$ but just  collections of $\delta$-functions.
}

 {
Anyway, we expect that
$\overline{\langle\sigma_z\rangle}_E=0$, unless there is a mechanism for
breaking the symmetry between components $A$ and $B$ of the dimer. We
assume it is excluded here, and the same would hold even if localization
occurs. Therefore, the quantity of interest is a more complicated
object, which is  the second moment
$\overline{\langle\sigma_z\rangle^2}_E$. It can be computed applying
the same trick as used in calculation of the inverse participation
ratio \cite{met_ner_bol_10,slanina_12b}. The core quantity is the
disorder-averaged product of resolvents taken at two
complex-conjugated points. If $\overline{R(\zeta_+)R(\zeta_-)}$ is
known, we can compute
$\mathcal{R}^{(2)}(\zeta_+,\zeta_-)=\sum_{IJij}(-1)^{I+J}\overline{
R_{IiIi}(\zeta_+)R_{JjJj}(\zeta_-)}$
and from here
\begin{equation}
\overline{\langle\sigma_z\rangle^2}_E=
\lim_{\epsilon\to 0^+}\frac{
\epsilon\mathcal{R}^{(2)}(E+\mathrm{i}\epsilon,E-\mathrm{i}\epsilon)
}{
\mathrm{Im}\overline{\mathrm{Tr}R(E-\mathrm{i}\epsilon)}
}\;
\label{eq-app-sig2-from-r2}
\end{equation}
The key point in proving this formula consists, as shown in
\cite{met_ner_bol_10}, in the fact that the double sum over
eigenvalues, which is present in the product $R(\zeta_+)R(\zeta_-)$, in
fact reduces to a single sum, because
only terms with both eigenvalues equal survive the limit $\epsilon\to 0$.
}

 {
The quantities $\overline{\mathrm{Tr}R(\zeta)}$ and
$\mathcal{R}^{(2)}(\zeta_+,\zeta_-)$ are calculated using the replica
trick. As a first step we define the replicated partition functions
\begin{widetext}
\begin{equation}
Z(\zeta;n)=\int
e^{-\frac{1}{2}\zeta\sum_{Ii}\sum_{a=1}^n(\phi_{Ii}^a)^2}
\overline{\exp\Big(\frac{1}{2}\sum_{IiJj}\sum_{a=1}^n\phi_{Ii}^a\,H_{IiJj}\,\phi_{Jj}^a\Big)}\,d[\phi_{Ii}^a]
\end{equation}
\begin{equation}
Z^{(2)}(\zeta_+,\zeta_-,\tilde{\zeta}_+,\tilde{\zeta}_-;n)=\int
e^{-\frac{1}{2}\sum_{s=\pm}\sum_{Ii}(\zeta_s+(-1)^I\tilde{\zeta}_s)\sum_{a=1}^n(\phi_{Iis}^a)^2}
\overline{\exp\Big(\frac{1}{2}\sum_{s=\pm}\sum_{IiJj}\sum_{a=1}^n\phi_{Iis}^a\,H_{IiJj}\,\phi_{Jjs}^a\Big)}\,d[\phi_{Iis}^a]
\end{equation}
\end{widetext}
Here  we use the abbreviations
$d[\phi_{Ii}^a]=\prod_{Ii}\prod_{a=1}^n d\phi_{Ii}^a$,
$d[\phi_{Iis}^a]=\prod_{Ii}\prod_{s=\pm}\prod_{a=1}^n
d\phi_{Iis}^a$. The density of states is then
\begin{equation}
\overline{\mathcal{D}}(E)=\frac{2}{\pi}\lim_{\epsilon\to 0^+}\mathrm{Im}
\Bigg(\frac{\partial}{\partial \zeta}\lim_{n\to 0}\frac{\partial}{\partial n}Z(\zeta;n)\Bigg)_{\zeta=E+\mathrm{i}\epsilon}
\end{equation}
and the averaged double resolvent
\begin{equation}
\mathcal{R}^{(2)}(\zeta_+,\zeta_-)=4\lim_{\tilde{\zeta}_\pm\to 0}
\frac{\partial^2}{\partial \tilde{\zeta}_+\partial \tilde{\zeta}_-}\lim_{n\to 0}\frac{\partial}{\partial n}Z^{(2)}\;.
\end{equation}
(We omitted the arguments of $Z^{(2)}$ to lighten the notation.)
Further procedure is standard \cite{kam_mez_99}. Let us first show how
it works for the density of states. In the $N\to\infty$
limit we apply the saddle point method, which leads to minimization of
the free energy
\begin{equation}
\mathcal{F}=\frac{1}{2}\sum_I \sum_{ab}(q^I_{ab})^2+\frac{1}{2}\ln\det M
\end{equation}
with respect to the two $n\times n$ matrices with elements $q^I_{ab}$.
Here
$M$ is square matrix $(2n)\times(2n)$ with elements
\begin{equation}
M_{IaJb}=(\zeta\delta_{ab}-\sqrt{2}\eta\,q^I_{ab})\delta_{IJ}
-d\,\delta_{ab}(1-\delta_{IJ})\;.
\end{equation}
Assuming replica symmetry
$q^I_{ab}=\overline{q}^I\delta_{ab}-\tilde{q}^I$ we obtain at the end
\begin{equation}
\overline{\mathcal{D}}(E)=
\frac{2}{\pi\,\eta}\lim_{\epsilon\to 0^+}\mathrm{Im}Q(E-\mathrm{i}\epsilon)
\end{equation}
where $Q(\zeta)$ is the solution of the cubic equation
\begin{equation}
Q=\frac{\frac{\zeta}{\eta}-Q}{
\big(\frac{\zeta}{\eta}-Q\big)^2-\big(\frac{d}{\eta}\big)^2}\;.
\end{equation}
In the course of the calculations it turns out that the off-diagonal
elements are all zero, $\tilde{q}^I=0$.
The same procedure can be used in calculation of
$\overline{\mathcal{D}_z}(E)$. The result is explicitly zero, as expected.
}

 {
Similar procedure is used in calculation of $Z^{(2)}$ and then
$\mathcal{R}^{(2)}$. Formally, it is nearly equal, with free energy
\begin{equation}
\mathcal{F}=\frac{1}{2}\sum_{Iss'} \sum_{ab} (q^{Iss'}_{ab})^2+\frac{1}{2}\ln\det M
\end{equation}
and  $(4n)\times(4n)$ matrix
\begin{equation}
\begin{split}
M_{IsaJs'b}&=\\((&\zeta_s+(-1)^I\tilde{\zeta}_s)\delta_{ss'}\delta_{ab}
-\sqrt{2}\eta\,q^{Iss'}_{ab})\delta_{IJ}\\
-&d\,\delta_{ss'}\delta_{ab}(1-\delta_{IJ})\;.
\end{split}
\end{equation}
Technically, however, it is more complicated,
because we have to minimize the free energy with respect to eight
$n\times n$
matrices $q^{Iss'}$. Even using the symmetry $s \leftrightarrow s'$ and assuming
replica-symmetric solution, leads to minimization with respect to six
variables $\overline{q}^{I++}$, $\overline{q}^{I+-}$,
$\overline{q}^{I--}$, where $I=A,B$. (As in the calculation of the
density of states, the off-diagonal elements of the
replica-symmetric matrices turn out to be zero explicitly). Let us
show the key steps of the replica-symmetric calculation. We can define
the $4\times 4$ matrix $m$ with elements
\begin{equation}
\begin{split}
m_{IsJs'}=((\zeta_s+(-1)^I\tilde{\zeta}_s)\delta_{ss'}
-\sqrt{2}\eta\,\overline{q}^{Iss'})\delta_{IJ}\\
-d\,\delta_{ss'}(1-\delta_{IJ})\;.
\end{split}
\end{equation}
We also define the auxiliary constant matrices with
\begin{equation}
T^{Iss'}=-\frac{1}{\sqrt{2}\,\eta}\frac{\partial\,m\,\,}{\partial\, \overline{q}^{Iss'}}
\end{equation}
and
\begin{equation}
V^s=\frac{\partial \,m}{\partial \,\tilde{\zeta}_s}\;.
\end{equation}
All matrix elements of these matrices are either $0$ or $1$ or $-1$
and they are useful for making the formulas compact.
Then, the set of equations for the parameters of the replica-symmetric
solution is
\begin{equation}
\overline{q}^{Iss'}=\frac{\eta}{\sqrt{2}}\mathrm{Tr}\,m^{-1}\,T^{Iss'}\;.
\end{equation}
This is a set of six independent algebraic equations of fifth
order. When solved, we can insert them to find $Z^{(2)}$ and by
derivation with respect to $\tilde{\zeta}_+$ and  $\tilde{\zeta}_-$ we
obtain the formula for $\mathcal{R}^{(2)}$. To this end, we fist need
the $8\times 8$ matrix $G$, whose inverse is found explicitly
\begin{equation}
\begin{split}
(G^{-1})_{Is_1s_2\,Js'_1s'_2}&=\\
\delta_{IJ}&\delta_{s_1s'_1}\delta_{s_2s'_2}+\\
\eta^2&\mathrm{Tr}\,T^{Is_1s_2}m^{-1}T^{Js'_1s'_2}m^{-1}\;.
\end{split}
\end{equation}
Finally, we obtain
\begin{widetext}
\begin{equation}
\begin{split}
\mathcal{R}^{(2)}(\zeta_+,\zeta_-)=&\,
2\mathrm{Tr}\,V^+m^{-1}V^-m^{-1}-\\
&-2\eta^2\,\sum_{Is_1s_2}\sum_{Js'_1s'_2}
\Big(\mathrm{Tr}\,V^+m^{-1}T^{Is_1s_2}m^{-1}\Big)
G_{Is_1s_2\,Js'_1s'_2}
\Big(\mathrm{Tr}\,V^-m^{-1}T^{Js'_1s'_2}m^{-1}\Big)
\end{split}
\end{equation}
\end{widetext}
where we assumed that the limit $\tilde{\zeta}_\pm\to 0$ was taken, so
the right-hand side is function of $\zeta_+$ and $\zeta_-$ only.
}

 {The formulas listed above solve the problem of
computing the second moment of the $z$-component of local
spin. However, it has non-zero value only on condition that
$\mathcal{R}^{(2)}(E+\mathrm{i}\epsilon,E-\mathrm{i}\epsilon)$
diverges for $\epsilon\to 0$. In the calculation of inverse
participation ratio this happens in the localized phase.
In our case, there is no source of
such divergence. Therefore, after performing the limit
$\epsilon\to 0$ in (\ref{eq-app-sig2-from-r2}) we get
$\overline{\langle\sigma_z\rangle^2}_E =0$. In fact, this is exactly
what was expected in the delocalized phase. However, this holds only
in the $N\to\infty$ limit and it would be highly desirable to
compute finite size effects, in order to compare it to the
simulation results. This is a considerably more involved
project. The first and most obvious source of finite-size
corrections is the fluctuation around the minimum if
$\mathcal{F}$. This can be obtained by calculating the Hessian
matrix at the replica-symmetric minimum and then performing the Gaussian
integration. However, there is also another possible source of
finite-size effects. Indeed, as showed in \cite{kam_mez_99}, in a
typical situation there
are also replica-symmetry-broken solutions with contribution of the
order $1/N$. And even worse, there is whole manifold of such
solutions and it is indispensable to integrate over entire such
manifold. We expect that the experience from  \cite{kam_mez_99} will
apply also in our case. We have no ambition of making such
calculations in this work, so we stop at this stage.}

 {Let us finish with a perspective to possible future
calculations. Here we used a dimer of coupled GOE matrices. It is
natural to extend the random-matrix ensemble further in order to
allow for localized states. One such choice is the Rosenzweig-Porter
ensemble, which was (besides the supersymmetric approach, see e.g.
\cite{kra_kha_cue_ami_15}) recently solved also by replica method
\cite{ven_cug_sch_tar_23}. So, we propose to make a dimer of two
Rosenzweig-Porter matrices and observe the localization transition
in the behavior of $\overline{\langle\sigma_z\rangle^2}_E$, as
sketched in the calculations above. However, one important
observation is due here. In many earlier works on Rosenzweig-Porter
matrices the transition is driven by the exponent determining the
scaling of matrix elements with system size $N$. On the contrary, in
our case the transition is driven by noise strength $\eta$. In fact,
this is just how the problem was formulated in the founding paper of
Rosenzweig and Porter \cite{ros_por_60}. It is the latter
formulation that is relevant for our purposes.}

\end{document}